\documentclass[11pt,a4paper,twoside]{paper}
\usepackage{amsmath,amssymb,amsthm}
\usepackage{graphicx}
\usepackage{parskip}
\usepackage{bm}
\usepackage{bbm} 
\usepackage{float}
\usepackage{xcolor}

\paragraphfont{\sffamily\itshape}

\newtheoremstyle{etoile}{\parskip}{\parskip}{\itshape}
                        {0pt}{\bfseries\sffamily}{.}{ }{}
\theoremstyle{etoile}

\makeatletter
\@addtoreset{equation}{section}
\makeatother

\newcommand\egaldef{\stackrel{\mbox{\upshape\tiny def}}{=}}

\newcommand\1{\leavevmode\hbox{\rm 
\small1\kern-0.35em\normalsize1}}
        
\newcommand\E{{\mathbb E}}
\newcommand\LL{{\mathcal L}}
\newcommand\D{{\mathcal D}}
\newcommand\F{{\mathcal F}}
\newcommand\I{{\mathbb I}}

\newcommand\n{^{\scriptscriptstyle (N)}}
\newcommand\nt{^{{\scriptscriptstyle (N)}t}}
\newcommand\nn{^{\scriptscriptstyle (N+1)}}
\newcommand\nnt{^{{\scriptscriptstyle (N+1)}t}}
\newcommand\sn{{\scriptscriptstyle N}}

\newcommand\bk{{\bf k}}
\newcommand\f{{\bf f}}
\newcommand\bu{{\bf u}}
\newcommand\bv{{\bf v}}
\newcommand\w{{\bf w}}
\newcommand\x{{\bf x}}

\newcommand\Tr{\text{Tr}}

\newcommand\R{\mathbb R}

\newcommand\N{\mathcal N}
\newcommand\gd{\mathcal Q}

\def\DD{\displaystyle} 

\DeclareMathOperator*{\Lra}{\longrightarrow}

\begin{document}

\title{Free Dynamics of Feature Learning Processes}

\author{Cyril Furtlehner 
\thanks{Inria Saclay - Tau team, LISN - AO team, B\^at 660 Universit\'e Paris-Saclay, Orsay Cedex 91405}}
\date{October 2022}

\maketitle

\abstract{
  Regression models usually tend to recover a noisy signal in the form of a combination of regressors, also called features in machine learning, themselves being the result of a learning process.  
  The alignment of the prior covariance feature matrix with the signal is known to play a
  key role in the generalization properties of the model, i.e. its ability to make predictions on unseen data during training. We present a statistical physics picture of the learning
  process. First we revisit the ridge regression to obtain compact asymptotic expressions for train and test errors, rendering manifest the conditions under which efficient generalization occurs.  
  It is established thanks to an exact test-train sample error ratio combined with random matrix properties.
  Along the way in the form of a self-energy emerges an effective ridge penalty \textemdash\ precisely the train to test error ratio \textemdash\ which offer a very simple parameterization of  the problem.
  This formulation appears convenient to tackle the learning process of the feature matrix itself. 
  We derive an autonomous dynamical system in terms of elementary degrees of freedom of the problem determining the evolution of the relative alignment between the population matrix and the signal.
  A macroscopic counterpart of these equations is also obtained and various dynamical mechanisms are unveiled, allowing one to interpret 
  the dynamics of simulated learning processes and reproduce trajectories of single experimental run with high precision.
}

\section{Introduction}
\begin{figure}
\centerline{\resizebox{0.8\textwidth}{!}{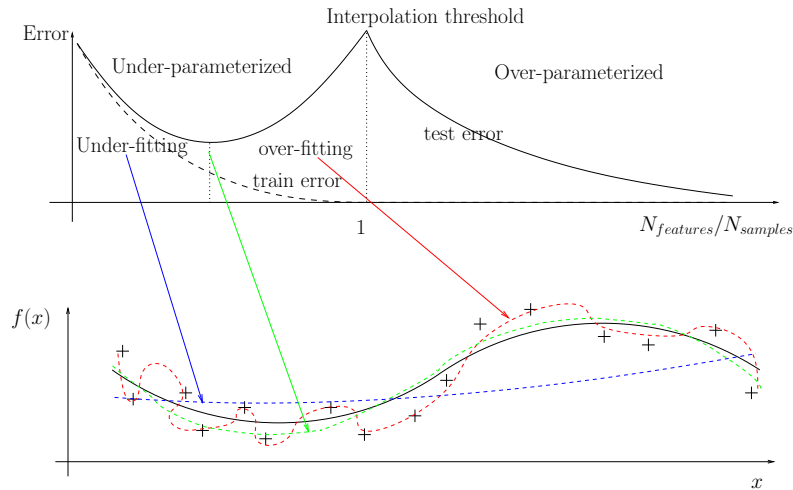}}
\caption{\label{fig:double_descent} Double descent ``paradox''. On the left is the traditional picture of a $U$-shape curve of the test error showing first under-fitting corresponding to high bias,
  followed by over-fitting characterized by high variance as illustrated on the $1$-d regression problem. The right part corresponds often to the deep learning setting with high complexity/expressivity
  of the model along with low test error.}
\end{figure}

The rapid developments of the last decades in machine learning (ML), in particular around the deep neural networks (DNN), have led statisticians to revisit
the mathematical foundations of the field~\cite{geman1992neural,Vapnik}, where at odds with intuition, highly over-parameterized models (typically DNN)
perform well in generalization~\cite{ZhBeHaReVi_2021,belkin2018understand}. In the context of supervised learning that we are interested in here, data are given in the form of input-output pairs $(\x,y)$
and the goal is to learn a function $f$ associating $\x$ to $y$. 
The input $\x\in\R^d$ belong to some $d$-dimensional embedding space, $d$ being possibly large and the $y\in\R$ is taken as a scalar for simplicity.
The function is often chosen in some parameterized class of function $f_\theta$, like neural networks, after a loss function $\LL(\theta\vert \D)$
has been designed and minimized conditionally to some training set $\D = \{(\x_i,y_i),i=1,\ldots N\}$.
Then its ability to generalize well on unseen data is assessed from  the test error, estimated on another set of data called the test set, statistically independent from the training set.
The essence of the dilemma has been summarized as the double descent picture in~\cite{belkin2019reconciling}, which is reproduced here in Figure~\ref{fig:double_descent}.
The traditional picture concerns the transition from under-fitting to over-fitting as the complexity of the model is increased corresponding to the best trade-off between bias and variance as
illustrated on the $1$-d regression model. At some (sweet) point, while the train error continues to decrease, the test error instead starts to increase again to reach a maximum
at the interpolation threshold where the number of parameters equals the number of train examples. The novelty with the advent of deep learning is that when going far into the over-parameterized region,
better generalization performances can sometime be obtained than those corresponding to the sweet spot separating the underfitting from the overfitting region.

This observation has revived the interest in the generalization mechanism in ML
and  triggered a vast number of works during the last few years, which we only give here a very brief and partial account. Most of them point to the effect of some form of implicit regularization,
meaning that either the architecture itself or the learning process itself provide some inductive bias penalizing automatically ``complex'' features
schematically corresponding to the UV spectrum, and favoring ``simpler'' one, i.e. typically in the IR part of the spectrum.
In a series of paper the linear ridge regression was revisited~\cite{dobriban2018high,AdSaSo_2020,BaLoLuTs,hastie2022surprises}, with various averaging hypothesis regarding the signal and the input data,
and showing to exhibit in some cases this double descent behaviour. In~\cite{AdSaSo_2020,liao2018dynamics},
is remarked in particular  that some implicit regularization is provided by the learning dynamics itself,
where the lower part of the spectrum of the feature matrix which is potentially harmful is also learned after the stronger hence justifying early stopping strategies, while some spectral gap
expected from random matrix theory may protect against overfitting in the overparameterized regime. 
In~\cite{BaLoLuTs} another mechanism of implicit regularization called benign over-fitting is unveiled.
It may occur when the population matrix has a large number of low eigenvalues decaying slowly. This is interpreted as self induced regularization~\cite{bartlett2021deep},
where the lower part of the prior spectrum, the ``tail'' features, acts as an $L_2$ regularization if the corresponding eigenvalues span a bounded interval.
Extending asymptotic results of~\cite{dobriban2018high} based on random matrix theory (RMT), another mechanism explaining good/bad generalization properties 
is identified in~\cite{richards2021asymptotics,wu2020optimal}. This mechanism involves both the signal and the population matrix, by considering the alignment between the two
and was already discussed in~\cite{louart2018random} in the context of NN with random projections.
Good generalization properties are observed whenever the signal decomposes preferentially on the strong modes of the population matrix. This in our point of view is a key point to
characterize feature quality and generalization, and we elaborate on this in this work.
These considerations may have also implications to define pre-conditioning strategies~\cite{amari2020does} in order to favor the modes mostly correlated  with the signal.
Extensions of double-descent analysis to non-linear models have been also explored in various works~\cite{ba2019generalization,mei2022generalization}. Concerning the conditions of use of RMT,
these are often  met in the context of ML to allow precise analysis~\cite{couillet2022random}, because ML models typically involve large weights matrices of weekly correlated entries. This also offer
the possibility to study the learning dynamics as already exemplified in~\cite{AdSaSo_2020,liao2018dynamics}. 
 
On the statistical physics side many works have also been discussing the double descent and generalization mechanisms. In~\cite{geiger2020scaling} it is argued that
large neural network perform as model ensembles through collective effects of roughly independent part of the NN, automatically reducing the variance which is otherwise expected
to dominate in the overparameterized regime. Methods inherited from the spin-glass theory~\cite{MePaVi} allow one in principle to study a great variety of learning scenario~\cite{seung1992statistical}
but need some efforts to go beyond oversimplified toy model architecture and data distributions to be of practical interest to ML. For quite some time now, many results have been obtained 
in this direction by various groups, concerning prediction for learning curves in various  non-trivial
regimes~\cite{malzahnOpper2002,malzahnOpper2005,loureiro2021learning,gerace2020generalisation,spigler2020asymptotic,CoMaRi_2021,rocks2022memorizing} and various settings around 
the teacher-student one~\cite{zdeborova2016statistical} for instance.
Another convenient setting to analyze the learning process is to consider the kernel regime or equivalently  the random feature model~\cite{rahimi2007random}. In these cases the data are sent respectively
into an infinite dimensional space or at least to a sufficiently large dimensional space where regression or classification tasks can be done efficiently with linear methods.
These regimes are found to be relevant when analyzing large NN when the width of the layers are very large. It is referred to as the lazy training regime~\cite{ChOyBa}, because there the weights
defining the features barely change after initialization. In this case 
the neural network behave as a Gaussian process~\cite{neal2012bayesian,williams1995gaussian} as can be proved in the limit of infinite size layers~\cite{lee2018deep}. It then
performs as kernel regression with a deterministic kernel called neural tangent kernel~\cite{JaGaHo_2018} resulting from central limit theorem applied layer-wise.

In this work we provide new insights to generalization mechanisms and feature learning, by exploiting asymptotic results obtained in the standard RMT regime, namely the planar diagram approximation,
and setting up a statistical physics description of the dynamics of the learning process. 
In order to make the paper self-contained, we first recall
in Section~\ref{sec:NN2RR} to which extent linear ridge regression is relevant to understand NN and discuss in Section~\ref{sec:bias-variance} the bias-variance dilemma in this context.
Then Section~\ref{sec:RR} is devoted to obtain general asymptotic expressions, in the context of  the ridge regression, for the train, test error and loss functions.
First we show in Section~\ref{sec:Test-train} the existence of a simple relation between train and test error for a given sample and how this is converted asymptotically to a deterministic train-test error
ratio in Section~\ref{sec:asymp}, while in Section~\ref{sec:LambdaEff}, the Feynman diagram formalism used to justify these formulas leads us also to define the effective ridge penalty of
the problem, which then considerably simplifies the analysis of the generalization error given at the end of this section.
In Section~\ref{sec:learningdynamics} we turn to the analysis of the learning dynamics intended to understand the learning process of the features themselves.
We consider a semi-lazy regime, the parameters of the features being in the lazy regime, while the weights of the last layer solve a ridge regression conditionally on the features themselves.  
In Section~\ref{sec:micro},
based on the asymptotic expressions obtained in Section~\ref{sec:RR}, we derive an autonomous set of dynamical equations
corresponding to the microscopic level description, i.e. concerning the dynamics of the basic degrees of freedom of the problem.
Macroscopic counterparts of these equations are given in Section~\ref{sec:macro} and special cases are studied in Section~\ref{sec:Specialcases2}.

\section{From neural networks to ridge regression}\label{sec:NN2RR}
\subsection{Linear regime of learning}\label{sec:ntk}
In some regimes neural networks can behave approximately as linear regression models~\cite{JaGaHo_2018}.
The starting point is to regress some signal observed  in the form  
\begin{equation}\label{eq:noisy_signal}
y =f(\x) + \epsilon
\end{equation}
where the signal itself corresponds to some unknown function $f$ on some input $\x\in\R^d$, with $d$ the supposedly large embedding dimension of the input,
sumperimposed with some decorrelated noise $\epsilon = \N(0,\sigma_\epsilon^2)$.  
Given some training data $\D = \{(\x^{(s)},y^{(s)}),s=1,\ldots N\}$, the regression problem amounts then to find the best minimizer $f_\theta$ among a parametric family indexed by $\theta$,
of the loss function chosen by convenience to be MSE:
\begin{equation}\label{eq:loss}
\LL(\theta) = \frac{1}{2N}\sum_{s=1}^N \vert y^{(s)}-f_\theta(\x^{(s)})\vert^2. 
\end{equation}
Assume first that we are given a supposedly large set of $N_f$ random features, $\{f_k(\x),k=1,\ldots N_f\}$, then our model can be chosen in the linear family
\begin{equation}\label{eq:lineareg}
f_\theta(\x) = \sum_{k=1}^{N_f} w_k f_k(\x),
\end{equation}
with $\theta = \{w_1,\ldots,w_{N_f}\}$. In order this to have a chance to work we need at least that the projection $f^\parallel$ of $f$ on the subspace spanned by these features
represents a significant fraction of the whole signal. If for instance we look for a smooth function with no privileged direction in space, requiring
a resolution $a$ on a domain of linear size $L$, we typically need a number of features
\[
D = \Bigl(\frac{L}{a}\Bigr)^d,
\]
which becomes prohibitive as soon as $d\gg 1$. The way to overcome the so-called ``curse of dimensionality'' is to consider non-random features obtained by some selection procedure. Multilayer neural
network for instance constitute an efficient solution to do that, where typically the last layer corresponds to~(\ref{eq:lineareg}) with $f_k = f_k^{(N_l)}$ and the recursive definition 
\[
f_k^{(n)}(\x) = \phi\Bigl[\sum_\ell w_{k\ell}^{(n-1)} f_\ell^{(n-1)}(\x)-b_k^{(n)}\Bigr] 
\]
of the feature of the $n$th layer, $\phi$ being the so-called activation function. The parameters of this family now correspond to $\theta = \{w_{k,\ell}^{(n)},b_k^{(n)}\}$ and the 
retro-propagation of the gradient of the loss~(\ref{eq:loss}) through the layer will lead in principle to find better than random features.

Still, as noticed in~\cite{JaGaHo_2018} a linear regime of learning can be recovered on this complex models. To see this
first we may consider the continuous limit of the learning process indexed by ``time'' $t$ 
\begin{equation}\label{eq:thetadot}
\partial_t\theta_t = -\gamma\nabla_\theta\LL(\theta_t),
\end{equation}
$\gamma$ representing the learning rate in this limit. In~\cite{JaGaHo_2018} is introduced the following quantity  
\begin{equation}\label{eq:NTK}
G_\theta(\x,\x') \egaldef \nabla_\theta^T f_\theta(\x)\nabla_\theta f_\theta(\x'),
\end{equation}
called ``neural tangent kernel'' (NTK) which as we shall see is the equivalent of a Green function in the context of field
theories, and can serve for instance as a measure of similarity between points as seen from a NN~\cite{charpiat2019input}.
This kernel allows one to study the learning dynamics directly in the  linear  space of functions $f: \R^D\longrightarrow \R$
according to the following equation
\begin{equation}\label{eq:NTK_dyn}
  \partial_t f_{\theta_t} = -\gamma G_{\theta_t} \hat P \bigl[f_{\theta_t}-f^\star\bigr],
\end{equation}
where
\[
\hat P(\x,\x') = \frac{1}{N}\sum_{i=1}^N \delta(\x-\x_i)\delta(\x'-\x_i),
\]
is the projection operator on the data and $f^\star$ is a function verifying
\[
f^\star(\x_i) = y_i,\qquad i=1,\ldots N.
\]
In the limit of infinitely large layers and with weights scaling like $w_{k\ell} = {\cal O}(1/\sqrt{N_l})$ if $N_l$ is the size of the layers
the NTK becomes a deterministic constant kernel $G_{\theta_0}$ yielding a linear dynamics in~(\ref{eq:NTK_dyn}). In this regime each component of $\bigl[f_{\theta_t}-f^\star\bigr]$
on the eigenmodes of $G_{\theta_0}$ evolves independently, and converges with some exponential decay at a speed determined by the corresponding eigenvalue.
As remarked in~\cite{AdSaSo_2020} the strategy called ``early stopping''  yields a form of regularization by avoiding the learning of the components of $f_{\theta_t}$ corresponding to low
eigenvalues, i.e. direction of the feature space which are poorly explored by the data as will be made more clear later on. This regime, where the parameters of the NN are displaced by the learning process
from their initial values by quantities scaling like $\mathcal{O}\Bigl(1/\sqrt{N_l}\Bigr)$, is also called ``lazy training" regime.
The interpretation of this is the following. Let $\theta_0$ be the parameter point at which $f_\theta$ is initialized at the beginning of the learning. Assuming small variations around $\theta_0$ 
we have at first order
\[
f_\theta(\x) = f_{\theta_0}(\x) + \sum_{k=1}^{N_f} \delta\theta_k\frac{\partial}{\partial \theta_k}f_\theta(\x)\Bigr\vert_{\theta=\theta_0}
\]
with $\delta\theta = \theta-\theta_0$ representing now the weights of the linear regression model defined in~(\ref{eq:lineareg}), after identifying the features 
\[
f_k(\x) = \frac{\partial}{\partial \theta_k}f_\theta(\x)\Bigr\vert_{\theta=\theta_0},
\]
where the input vector is now $(f_k(\x_i),k=1,\ldots N_f)$ of dimension $N_f$, for a given input data $\x_i$.
As discussed above, for large dimensional problem it is not clear that the features defined at initialization in this way have a good chance to be adapted to the problem,
hence this regime, even if interesting, is likely not able to explain alone the success of neural network on difficult tasks~\cite{ChOyBa}.

\subsection{The Bias-Variance trade-off for the ridge regression}\label{sec:bias-variance}
A fundamental aspect of learning with a limited number of data is the problem of choosing properly the complexity of the model given the size of the
training set of examples. This problem is formalized in terms of the bias-variance dilemma~\cite{geman1992neural} when the loss is a mean-squared error as in~(\ref{eq:loss}). 
Indeed in that case the test error can be decompose into bias and variance as follows. 
The optimal solution is formally given by 
\[
f(\x) = \E[y\vert\x]. 
\]
Then given a predictor $f(\x\vert\D)$ obtained from a specific training set, we have when averaging over $\D$ the
bias and variance decomposition of the test error: 
\[
\E_\D\Bigl[\bigl(f(\x\vert\D)-\E[y\vert\x]\bigr)^2\Bigr] = \Bigl(\E_\D[f(\x\vert\D)]-\E[y\vert\x]\Bigr)^2 + \E_\D\Bigl[\Bigl(f(\x\vert\D)-\E_\D[f(\x\vert\D)]\Bigr)^2\Bigr].
\]
Consistency of the predictor is realized when both bias and variance go to zero along with the size of the dataset $\D$ going to infinity.
Choosing the right complexity of the model at finite $N$ is therefore equivalent, at least in the traditional ML view,
as illustrated on Figure~\ref{fig:double_descent},
to obtain the best trade-off between bias and variance.
For model with low complexity we expect a high bias due to the lack  of expressivity of the model, but a low variance as there are few parameters to estimate compared to 
the number of training samples. Instead when the model has high complexity, it will be able to fit precisely  a noisy observation of the signal~(\ref{eq:noisy_signal}),
so that when averaging over $\D$ the bias get small, but the variance is expected to be larger since it cumulates the  noise of the signal with the noise of the model.

Let us illustrate this on the ridge regression, in order also to introduce some notations to be used later on.  
Consider the noisy observation  $y\in\R$  
\begin{equation}\label{eq:teacher}
y = \f^t\x +\epsilon,
\end{equation}
of some signal defined by the inner product of some fixed unknown vector $\f\in\R^{D}$ with some known input vector $\x\in\R^D$ of basic features 
and the noise $\epsilon$ of variance $\sigma^2$. The statistics by which $\x$ is represented in the data  
is summarized by the so-called ``population matrix''  in the statistics literature, i.e. the prior cross-product  of the  basic features, that we denote here by $C$,
\begin{equation}\label{eq:pop}
C \egaldef {\mathbb E}_\x [\x\x^t].
\end{equation}
The ridge regression problem amounts to find $\hat\w\in\R^D$ minimizing the loss
\begin{equation}\label{eq:ridge_loss}
\LL(\w) = \frac{1}{2}\w^t \w+\frac{\alpha}{2N}\sum_{s=1}^{N}\bigl\vert y^{(s)}-\w^t\x^{(s)}\bigr\vert^2
\end{equation}
where $\D = \{(\x^{(s)},y^{(s)}),s=1,\ldots N\}$ is a training set generated by the teacher model~(\ref{eq:teacher})
and $\alpha$ corresponds here to the inverse of the ridge penalty, introduced this way by convenience
and to be later interpreted as a bare coupling constant. $\rho = N/N_f$ is the aspect ratio of the data matrix. 
Given the empirical averages 
\begin{align*}
  C\n &\egaldef \frac{1}{N}\sum_{s=1}^N \x^{(s)} {\x^{(s)}}^t, \\[0.2cm]
  Z\n &\egaldef   \frac{1}{N}\sum_{s=1}^N \x^{(s)} y^{(s)},
\end{align*}
and the resolvent 
\begin{equation}\label{eq:Gn}
 G_\alpha\n = \bigl(\I+\alpha C\n\bigr)^{-1},
\end{equation}
with $\I$ being the identity matrix,
we obtain the optimal solution 
\begin{equation}\label{eq:ridge_sol}
\hat\w_\alpha\n = \alpha G_\alpha\n Z\n.
\end{equation}
From this we can deduce the following expressions from the train and test errors for a given dataset
\begin{align}
  E_{\rm train}\n(\alpha) &= \frac{1}{\alpha}\Tr\Bigl[G_\alpha\n(\I-G_\alpha\n)\f\f^t\Bigr]
  +\sigma^2\Bigl(1-\rho^{-1}+\frac{1}{N}\Tr\Bigl[{G_\alpha\n}^2\Bigr]\Bigr),\label{eq:E_train0}\\[0.2cm]
  E_{\rm test}\n(\alpha) &= \Tr\Bigl[G_\alpha\n C G_\alpha\n \f\f^t\Bigr]
  +\sigma^2\Bigl(1+\frac{\alpha}{N}\Tr\Bigl[G_\alpha\n(\I-G_\alpha\n) C\Bigr]\Bigr).\label{eq:E_test0}
\end{align}
after averaging over the noise and the test set following~(\ref{eq:pop}).
This last equation exhibits the bias-variance decomposition of the generalization error,
the first term on the rhs corresponding to the bias and the second one proportional to $\sigma^2$ corresponding to the variance. The effect of the coupling $\alpha$ is
to sharpen the model, by reducing the influence of the regularization term in~(\ref{eq:loss}) and the bias-variance dilemma becomes manifest by looking at the role of $\alpha$.
As we see, the variance term is composed of the contribution $\sigma^2$ due to the noise $\epsilon$ of the signal itself and a second term coming from the
fact that this noise is fitted by the model. This latter contribution can be harmful when $C^{(\sn)}$ has a high density of low eigenvalues especially when $\alpha$
is large and the way to contain it is to reduce $\alpha$.
Instead, the bias term which is also problematic when $C^{(\sn)}$ has a high density of low eigenvalues
can be reduced only by increasing $\alpha$.

For now we assumed that the model has all the necessary features at disposal to reconstruct the signal which is not
the standard case in practice. To take this into account we will in the following consider as in~\cite{AdSaSo_2020}  a slightly more general setting: the dimension $D$ of the embedding space of the
input $\x\in\R^D$ will be assume to be large, while the model is given a set of $N_f<D$ features in the form of an $N_f\times D$ matrix $F$, and we look now for a student model 
of the form
\begin{equation}\label{eq:mispecified}
y = \w^t F\x
\end{equation}
with now $\w\in\R^{N_f}$.
The population matrix has now some structure and reads
\begin{equation}\label{eq:popmatrix}
C = {\mathbb E}_\x\Bigl[F\x\x^t F^t\Bigr] = FF^t,
\end{equation}
assuming without loss of generality, i.e. up to a linear change of input vector, that $\x$ follows an isotropic and normalized distribution,
with ${\mathbb E}(\Vert \x\Vert^2) = D$.
The optimal solution is still given by~(\ref{eq:ridge_sol}) but with now
\begin{align}
  C\n &\egaldef \frac{1}{N}\sum_{s=1}^N F\x^{(s)} {\x^{(s)}}^tF^t, \label{eq:Cn}\\[0.2cm]
  Z\n &\egaldef   \frac{1}{N}\sum_{s=1}^N F\x^{(s)} y^{(s)}.\label{eq:Zn}
\end{align}

In addition $\f\f^t$ is replaced by $F^{+t}\f\f^t F^+$ in the expression of (\ref{eq:E_train0},\ref{eq:E_test0}) of the train and test errors, where $F^+$ denotes the pseudo-inverse
of $F$. Since $F$ has a rank smaller than $D$ we decompose the input in $\x= \x^\parallel+\x^\perp$, where $\x^\perp$ is the component of $\x$
in the kernel of $F$ s.t. $F\x = F\x^\parallel$. The part $\f^t\x^\perp$ of the signal cannot be recovered by the model which is now misspecified.
As discussed in~\cite{AdSaSo_2020}, this part can actually be considered as an additional contribution to the noise $\epsilon$ as soon as $\x^\perp$
is decorrelated from $\x^\parallel$, the noise variance being now changed to
\begin{equation}\label{eq:sigma_eff}
\sigma_{\rm eff}^2 \egaldef \sigma^2+{\mathbb E}\bigl[\Vert \f^t\x^\perp\Vert^2\bigr].
\end{equation}
This may change the traditional picture of a U-shape curve when increasing the number of parameters in the model and be responsible in some cases,
by decreasing the effective noise level as some particular rate, for the double descent
scenario observed in~\cite{belkin2019reconciling}. 
In summary the train and test error of the misspecified ridge regression problem are given by 
\begin{align}
  E_{\rm train}\n(\alpha) &= \frac{1}{\alpha}\Tr\Bigl[G_\alpha\n(\I-G_\alpha\n)F^{+t}\f\f^t F^+\Bigr]
  +\sigma_{\rm eff}^2\Bigl(1-\rho^{-1}+\frac{1}{N}\Tr\Bigl[{G_\alpha\n}^2\Bigr]\Bigr),\label{eq:E_train}\\[0.2cm]
  E_{\rm test}\n(\alpha) &= \Tr\Bigl[G_\alpha\n C G_\alpha\n F^{+t}\f\f^t F^+\Bigr]
  +\sigma_{\rm eff}^2\Bigl(1+\frac{\alpha}{N}\Tr\Bigl[G_\alpha\n(\I-G_\alpha\n) C\Bigr]\Bigr).\label{eq:E_test}
\end{align}

\subsection{Kernel learning, Gaussian processes  and Kernel regime}
In order to avoid the reader to get lost with the ML jargon already used at some places in particular in the introduction, we need to introduce 
closely related learning schemes and models which lead in the end to very similar analysis as the one developed for the ridge regression. 
The first one is kernel learning (see e.g.~\cite{bishop2006pattern}).
Based on a set of observations $\{\x^{(s)},y^{(s)},s=1,\ldots N\}$ it consists to define a prediction model for unseen data $y = f(\x)$ as a combination of training set observations:
\begin{equation}\label{eq:fK}
f(\x) = \sum_{s=1}^N K\bigl(\x,\x^{(s)}\bigr)y^{(s)},
\end{equation}
where $K$ is a well choosen kernel. Then to make connection with the linear regression model considered previously, consider the case where $f$ is sought in the form
of a linear combination of $N_f$ basic functions $\{\phi_k,k=1,\ldots N_f\}$, namely the features functions:
\[
f(\x) = \sum_{k=1}^{N_f} w_k\phi_k(\x), 
\]
where $\w$ is as before the vector of regression coefficients. The solution~(\ref{eq:ridge_sol}) can be made in correspondence with the kernel regression by simply identifying $K$ with
\begin{equation}\label{eq:K}
K(\x,\x') = \frac{\alpha}{N}\Phi^t(\x)\Bigl[\I+\frac{\alpha}{N}\sum_{s=1}^N \Phi\bigl(\x^{(s)}\bigr)\Phi^t\bigl(\x^{(s)}\bigr)\Bigr]^{-1}\Phi(\x')
\end{equation}
where $\Phi(\x) = (\phi_k(\x),k=1,\ldots N_f)$ is the feature vector.
This solution can be obtained as the mean of a Gaussian process which corresponds to a special case of a probabilistic, Bayesian approach to learning. 
In this setting $f$ is a random function which distribution called posterior, is given by
\[
P[f] = \frac{1}{Z} P_{\rm prior}[f] e^{-N\LL[f]} 
\]
where $P_{\rm prior}$ is the prior distribution, which happens to be Gaussian for Gaussian processes, $\LL[f]$ is the loss term like e.g.~(\ref{eq:loss}) attached to the training data,
while $Z$ is the normalization constant. Hence if $f$ is assumed to decompose on the set of feature $\phi_k(\x)$ previously considered, a Gaussian prior distribution on the
weights $\w$ would therefore be equivalent as to solving the ridge regression. Gaussian processes when considered as a Gaussian field theory can be analyzed with the ordinary tools of field
theory~\cite{bialek1996field,malzahnOpper2002,CoMaRi_2021}. For instance, the NTK mentioned in Section~\ref{sec:ntk} appears to be the $2$-points function
of the prior distribution as can be verified by identifying features $\phi_k(\x)$ as
\[
\phi_k(\x) \egaldef \frac{\partial}{\partial\theta_k}f_\theta(\x)
\]
and the adequate Gaussian field theory having $f$ specified by~(\ref{eq:fK},\ref{eq:K}) as the ground state.

\section{Spectral decomposition of ridge regression errors}\label{sec:RR}
\subsection{Test-train error ratio for the ridge regression}\label{sec:Test-train}
At least for simple machine learning models it is often possible to obtain relation between the train and test error,
which for instance leads to the notion of stability explored in~\cite{bousquet2002stability}, by bounding the difference between the test and train error typically.
Here on the specific case of the ridge regression, we obtain a different type  of relation, namely a ratio,  between test and train error, first when evaluated on a given sample.
The idea is to compare the solution~(\ref{eq:ridge_sol}) when the considered sample is incorporated or not in the train set. 

We will consider from now on the misspecified model~(\ref{eq:mispecified}). 
The solution obtained with $N$ training samples is given by~(\ref{eq:ridge_sol}). 
When one training data point is added we have the following recurrence:
\begin{align*}
Z\nn &= \frac{N}{N+1} Z\n+ \frac{1}{N+1} F\x\nn y\nn  \\[0.2cm]
G_\alpha\nn &= G_{\alpha'}\n-\gamma\n(\alpha')G_{\alpha'}\n F{\x\nn}{\x\nn}^t F^t G_{\alpha'}\n,
\end{align*}
thanks to the Sherman-Morrison formula,
with $\alpha' = \frac{N}{N+1}\alpha$ and
\[
\gamma\n(\alpha) \egaldef \frac{\alpha}{N+\alpha{\x\nn}^t F^tG_\alpha\n F\x\nn}.
\]
As a result we can compare $\hat\w_\alpha\n$ with $\hat\w_\alpha\nn$. Taking the transpose, we have
\[
{\hat\w_\alpha\nnt} = {\hat\w_{\alpha'}\nt}
+ \frac{\alpha' \x\nnt F^tG_{\alpha'}\n}{N+\alpha'{\x\nnt} F^tG_{\alpha'}\n F\x\nn}\bigl[y\nn-{\hat\w_{\alpha'}\nt} F\x\nn\bigr].
\]
Multiplying right by $F\x\nn$ and substracting  $\f^t\x\nn$ on both side of the equation we finally end up with the relation
\begin{equation}\label{eq:Errelation}
{\hat\w_{\alpha'}\nt} F\x\nn- y\nn = {\mathcal R}\n(\alpha',\x\nn)\bigl[{\hat\w_{\alpha}\nnt} F\x\nn-y\nn\bigr]
\end{equation}
with
\[
{\mathcal R}\n(\alpha,\x) \egaldef 1+\frac{\alpha}{N}{\x}^t F^t G_{\alpha}\n F\x.
\]
This quantity, greater than one, represents a test to train absolute error ratio, since on the l.h.s of~(\ref{eq:Errelation}) the difference of the model with the observation is evaluated
on $\x\nn$ not present in the train set while on the r.h.s $\x\nn$ is used to train the model $\hat\w\nn$.
In term of squared error the sample wise relationship between the test and train reads
\begin{equation}\label{eq:sampleTrTe}
  E_{\rm test}\n(\alpha',\x) = {{\mathcal R}\n}(\alpha',\x)^2 E_{\rm train}\n(\alpha,\x).
\end{equation}
which has to be average w.r.t. $\x$ to get the generalization error
\[
E_{\rm test}\n = {\mathbb E}_\x\Bigl[E_{\rm test}\n(\alpha,\x) \Bigr],
\]
Incidentally we found a similar statement made already some time ago in~\cite{malzahnOpper2005} in the context of Gaussian processes,
resulting however from a quite remote approach based on variational approximation to Bayesian learning.
As we shall see, ${{\mathcal R}\n}(\alpha,\x)$ for large $N$ is actually a self-averaging quantity, and since $\alpha$ and $\alpha'$ differ only by a ${\mathcal O}(1/N)$
quantity, it constitutes the actual test to train error ratio. 

\subsection{Asymptotic limits}\label{sec:asymp}
Equations~(\ref{eq:E_train},\ref{eq:E_test}) have been analyzed in the asymptotic limits by various authors
recently~\cite{dobriban2018high,wu2020optimal,richards2021asymptotics,hastie2022surprises}, using RMT when $N,N_f\to\infty$ with
fixed $\rho  = N/N_f$. Let us first discuss informally these results. The main difficulty resides in the first term of the l.h.s of  equation~(\ref{eq:E_test}) which involves three matrices,
the resolvent $G\n$ (from now on we drop in the notation the explicit dependency on $\alpha$),
the population matrix $C$ and the representation of the signal matrix $\tilde \f\tilde \f^t \egaldef F^+\tilde\f\tilde\f^tF^{+t}$ in feature space, with non mutually aligned sets of eigenvectors.
For instance, given the spectral density $\nu_\infty$ associated to $C$ and $h$ an arbitrary function, 
results of Ledoit-P\'echet~\cite{ledoit2011eigenvectors} allows one to get the following asymptotic limits:
\[
\lim_{N,N_f\to\infty\atop N/N_f=\rho} \frac{1}{N_f}\Tr\Bigl[G\n h(C) \Bigr] = \int dx \frac{\nu_\infty(x)h(x)}{1+\alpha x\bigl[1-\rho^{-1}+\rho^{-1}g(\rho)\bigr]}
\]
with
\[
g(\rho) \egaldef \lim_{N,N_f\to\infty\atop N/N_f=\rho} \frac{1}{N_f}\Tr\Bigl[G\n\Bigr],
\]
solution to the basic self-consistent equation of Marchenko-Pastur~\cite{MaPa}
\[
g(\rho) = \int dx \frac{\nu_\infty(x)}{1+\alpha x\bigl[1-\rho^{-1}+\rho^{-1}g(\rho)\bigr]},
\]
assuming no specific form of the population matrix. Instead, the combined term of interest in equation~(\ref{eq:E_test}) cannot be obtained in general by
the Ledoit-P\'echet formula, except for some special cases as in~\cite{dobriban2018high} where $\tilde\f\tilde\f^t$ is assumed isotropic or as in~\cite{wu2020optimal,richards2021asymptotics}
where $\tilde\f\tilde\f^t$ is assumed to be diagonal on the bases of $C$ eigenvectors, while in~\cite{hastie2022surprises} non-asymptotic expression of the terms in~(\ref{eq:E_train})
are obtained with weaker assumptions. In all these works, the expressions which are given are quite involved and difficult to interpret from the physics point of view.
Here we propose to take advantage of the relation~(\ref{eq:sampleTrTe}) and argue that it can be transposed at the asymptotic level in the form
\begin{equation}\label{eq:train-test}
E_{\rm test}(\rho,\alpha) = {\mathcal R}^2(\rho,\alpha) E_{\rm train}(\rho,\alpha)
\end{equation}
with
\begin{align*}
  E_{\rm test,train}(\rho,\alpha) &\egaldef \lim_{N,N_f\to\infty\atop N/N_f=\rho}{\mathbb E}_\x\Bigl[E_{\rm test,train}\n(\alpha,\x)\Bigr] \\[0.2cm]
  {\mathcal R}(\rho,\alpha) &\egaldef \lim_{N,N_f\to\infty\atop N/N_f=\rho}{\mathbb E}_\x\Bigl[{\mathcal R}\n(\alpha,\x)\Bigr]
  = 1+\lim_{N,N_f\to\infty\atop N/N_f=\rho}\Bigl[\frac{\alpha}{N}\Tr\Bigl(G_\alpha\n C\Bigr)\Bigr].
\end{align*}
The argument here is already used in the derivation of the Marchenko-Pastur distribution (see Appendix~\ref{app:diag_expansion}), namely that some traces are self-averaging when $N\to\infty$. 
Consider the $D\times D$ matrix $F^t G_{\alpha}\n F$, denote by $\{\lambda_k,k=1,\ldots N_f\}$ its non-zero eigenvalues assumed to be bounded and $\{x_k,k=1,\ldots D\}$ the components of
$\x$ on these modes. We have 
\[
\frac{1}{N_f}{\x}^t F^t G_{\alpha}\n F\x = \frac{1}{N_f}\sum_{k=1}^{D} x_k^2\lambda_k = \frac{1}{N_f}\Tr\Bigl[G_\alpha\n C\Bigr]+ {\mathcal O}\Bigl(\frac{1}{N_f}\Bigr),
\]
by law of large numbers, given that the $x_k$ are iid with $x_k = {\mathcal N}(0,1)$.
At this point we can directly use the Ledoit-P\'echet formula in a similar fashion to what is done in~\cite{dobriban2018high}
to express the asymptotic limits of ${\mathcal R}(\rho,\alpha)$ and $E_{\rm train}(\rho,\alpha)$ thereby giving $E_{\rm test}(\rho,\alpha)$ in~(\ref{eq:train-test}). As
explained in Appendix~\ref{app:diag_expansion} this can also be obtained thanks to a diagrammatic expansion restricted to a summation over non-crossing diagrams,
which corresponds to the combinatorial interpretation of free probabilities~\cite{voiculescu1991limit,mingo2017free},
which somehow backs all these asymptotic expressions of random matrix theory. The usefulness of the diagrammatic expansion  formalism
will be made more obvious in the next Section when  interpreting  the results.  
Eventually we arrive at the following spectral decomposition of the train error
and the test-train error ratio (see Appendix~\ref{app:TrainTestproof}):
\begin{align}
  {\mathcal R}(\rho,\alpha) &= 1+\frac{\alpha}{\rho}\int dx \nu_\infty(x) x g(x,\rho,\alpha),\label{eq:Rasymp}\\[0.2cm]
  E_{\rm train}(\rho,\alpha) &= -\int \frac{dx}{x} \mu^{\parallel}(x)\frac{\partial}{\partial \alpha}g(x,\rho,\alpha)
  +\frac{\sigma^2+\mu^\perp}{\rho}\Bigl(\rho-1+\frac{\partial}{\partial \alpha}\bigl[\alpha g(\rho,\alpha)\bigr]\Bigr).\label{eq:Etrainasymp}
\end{align}
with 
\begin{align*}
  g(\rho,\alpha) &= \int dx\nu_\infty(x) g(x,\rho,\alpha),\\[0.2cm]
g(x,\rho,\alpha) &= \left(1 + \frac{\alpha x}{{\mathcal R}(\rho,\alpha)}\right)^{-1}.
\end{align*}
Here, in addition to $\nu_\infty$, the spectral density of the population matrix $C$ hence associated to the features,
appear also spectral quantities associated to the signal $\mu^\parallel(x)$ and $\mu^\perp$.
Let us define them properly in thermodynamic limits.
Consider first the singular value decomposition (SVD) of the feature matrix:
\begin{equation}\label{eq:Fsvd}
F = \sum_{a=1}^{N_f} \sqrt{c_a}\bu_a\bv_a^t
\end{equation}
where the right singular vectors $\{\bv_a,a=1,\ldots D\}$ are arbitrarily completed  for $a=N_f,\ldots D$ to span the entire embedding space. 
The population matrix $C$ reads then
\[
C = \sum_{a=1}^{N_f} c_a \bu_a \bu_a^t,
\]
where $\{\bu_a,a=1,\ldots N_f\}$ and $\{c_a,a=1,\ldots N_f\}$ are the eigenvectors and eigenvalues of $C$ at finite $N_f$.
In thermodynamic limit we have
\[
\nu_\infty(x) \egaldef \lim_{N_f\to\infty} \frac{1}{N_f}\sum_{a=1}^{N_f}\delta(x-c_a).
\]
The signal itself is represented by  $\f$ defined on the embedding space $\R^D$.
On this bases we have the decomposition
\[
\f = \f^\parallel+\f^\perp = \sum_{a=1}^{N_f} f_a\bv_a + \sum_{a=N_f+1}^{D} f_a\bv_a.
\]
Then 
\begin{equation}\label{def:mupara}
\mu^\parallel(x) \egaldef \lim_{N_f\to\infty} \sum_{a=1}^{N_f} f_a^2\delta(x-c_a).
\end{equation}
It is normalized to 
\[
\mu^\parallel \egaldef \sum_{a=1}^{N_f} f_a^2,
\]
while
\[
\mu^\perp \egaldef \sum_{a=N_f+1}^D f_a^2
\]
represents the squared norm of the orthogonal part of the signal to the features.
The effective variance of the  noise defined previously in~(\ref{eq:sigma_eff}) reads in these notations $\sigma_{\rm eff}^2 = \sigma^2+\mu^\perp$.

\subsection{Effective coupling constant and spectral parameters}\label{sec:LambdaEff}
In the curse of derivation of these equations via the diagrammatic expansion, we obtain a closed form expression of the self energy involved in the Dyson equation
thanks to planar diagram approximation (see Appendix~\ref{app:diag_expansion}). 
In itself the self-energy, also called ${\mathcal R}$-transform in free probabilities,
represents a kind of dressing of the coupling constant $\alpha$ in the present expansion which suggests to define the following effective coupling
constant (see Appendix~\ref{app:TrainTestproof})
\begin{equation}\label{def:Lambdaeff}
\Lambda(\rho,\alpha) \egaldef \frac{\alpha}{{\mathcal R}(\rho,\alpha)}\le \alpha,
\end{equation}
as the inverse of the test-train ratio. Incidentally we found out in~\cite{Liu2020Ridge} that this quantity has a counterpart called the  
resolvent bias factor in the probabilistic/statistic literature. We found also a similar definition in~\cite{jacot2020implicit}
of an effective ridge penalty in the context of kernel regression, but resulting from different considerations.
Let us introduce two sets of statistical parameters: 
\begin{align}
  g_k[\Lambda] &\egaldef  \int dx\ \frac{\nu_\infty(x)}{[1+\Lambda x]^k},\label{def:g_k}\\[0.2cm]
  \mu_k[\Lambda] &\egaldef \frac{1}{\mu^\parallel}\int dx \frac{\mu^\parallel(x)}{\bigl[1+\Lambda x\bigr]^k}, \label{def:mu_k} 
\end{align}  
the former being associated to the feature matrix while the latter is associated to the signal. These will be of use later on. They represent
the moments of the variable $y=1/(1+\Lambda x)$ when $x$ is respectively drawn from $\nu_\infty$ and from $\mu^\parallel(x)/\mu^\parallel$. For the moment
let us present some properties concerning $\Lambda(\rho,\alpha) = \Lambda$ (the explicit dependency of $\Lambda$ w.r.t. $\alpha$ and $\rho$
will be omitted from  now on).
From~(\ref{eq:Rasymp}) we see that $\Lambda$ is solution of the self-consistent equation
\begin{align}
  \Lambda &= \alpha\bigl(1-\frac{1}{\rho}\bigr) + \frac{\alpha}{\rho}\int dx\frac{\nu_\infty(x)}{1+\Lambda x}\nonumber \\[0.2cm]
  &= \alpha\bigl(1-\frac{1}{\rho}\bigr) + \frac{\alpha}{\rho} g[\Lambda] \egaldef h_{\rho,\alpha}(\Lambda), \label{eq:FP}
\end{align}
where $g$ ($=g_1$) now reads in terms of this effective coupling, 
\[
g[\Lambda] = \int dx\ \frac{\nu_\infty(x)}{1+\Lambda x},
\]
therefore $h_{\rho,\alpha}$ is a strictly decreasing convex function of $\Lambda$ with $h_{\rho,\alpha}(0) = \alpha >0$.
As a result equation~(\ref{eq:FP})  admits a single solution
$\Lambda(\rho,\alpha) \in [0,\Lambda_{\rm max}(\rho,\alpha)[$ with
\begin{equation}\label{eq:gmin}
\Lambda_{\rm max}(\rho,\alpha) =
\begin{cases}
  \alpha\hspace{3.3cm} \rho>1, \\[0.2cm]
  \min(\alpha,g^{-1}[1-\rho]) \qquad\rho\le 1.
\end{cases}  
\end{equation}
In practice this solution can be rapidly obtained by iteration of $h_{\rho,\alpha}(\Lambda)$ or by dichotomic search in absence of convergence of the fixed point equation. 
Notice also that $g[\Lambda]$ is strictly decreasing, hence $g[\Lambda]\in]g_{\rm min}(\rho),1]$ with
\[
g_{\rm min}(\rho) \egaldef \max(0,1-\rho)
\]
Using the fixed point equation~(\ref{eq:FP}) we get the following expressions for the derivative of $\Lambda$ w.r.t. $\alpha$:
\begin{equation}
  \frac{\partial}{\partial\alpha}\Lambda(\rho,\alpha) = \frac{\Lambda^2}{\alpha^2}\frac{\rho}{\rho-\gd\bigl[\Lambda\bigr]},\label{eq:dLambda}
\end{equation}
with
\begin{align}
  \gd[\Lambda] &\egaldef \int dx\ \nu_\infty(x)\frac{\Lambda^2 x^2}{\bigl[1+\Lambda x\bigr]^2},\label{def:QLambda}\\[0.2cm]
    &= 1-2g[\Lambda]+g_2[\Lambda],\label{eq:QLambda}\\[0.2cm]
    &= \rho+\Lambda\Bigl(g'[\Lambda]-\frac{\rho}{\alpha}\Bigr).\nonumber
\end{align}
This last equality combined to~(\ref{eq:FP}) allows one to rewrite~(\ref{eq:dLambda}) in the form
\begin{equation}
\frac{\partial}{\partial\alpha}\Lambda(\rho,\alpha) = \frac{\Lambda}{\alpha}\frac{\rho}{\rho-\alpha g'\bigl[\Lambda\bigr]} > 0 \label{eq:dLambda2},
\end{equation}
the inequality following the fact that $g'[\Lambda]<0$.
Hence we get here the important property that the effective coupling $\Lambda$ is a strictly increasing function of the bare coupling $\alpha$, meaning that $\Lambda$
can replace $\alpha$ to parameterize the ridge regression.  
\begin{figure}
\centerline{\includegraphics[width=0.7\textwidth]{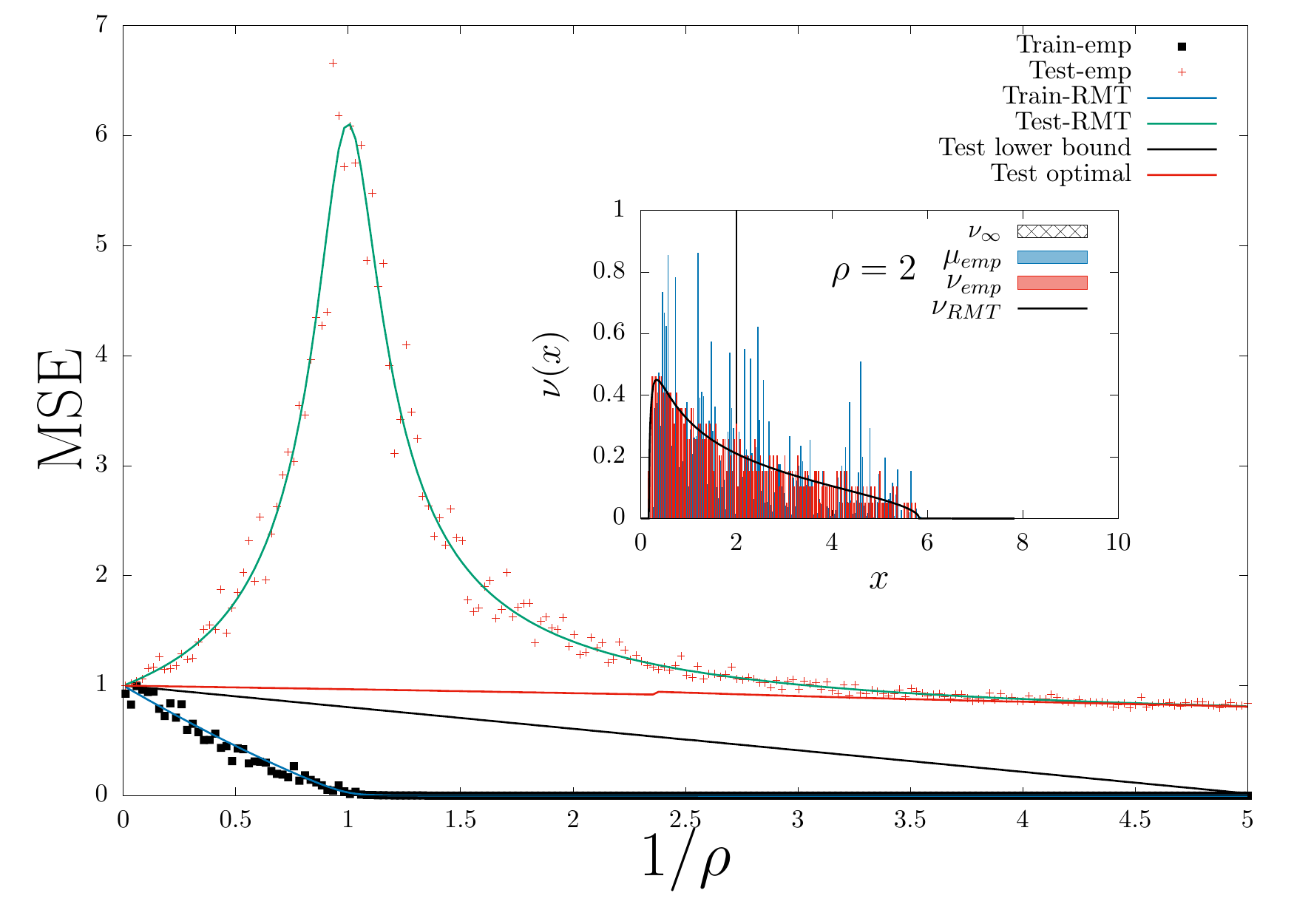}}
\centerline{\includegraphics[width=0.7\textwidth]{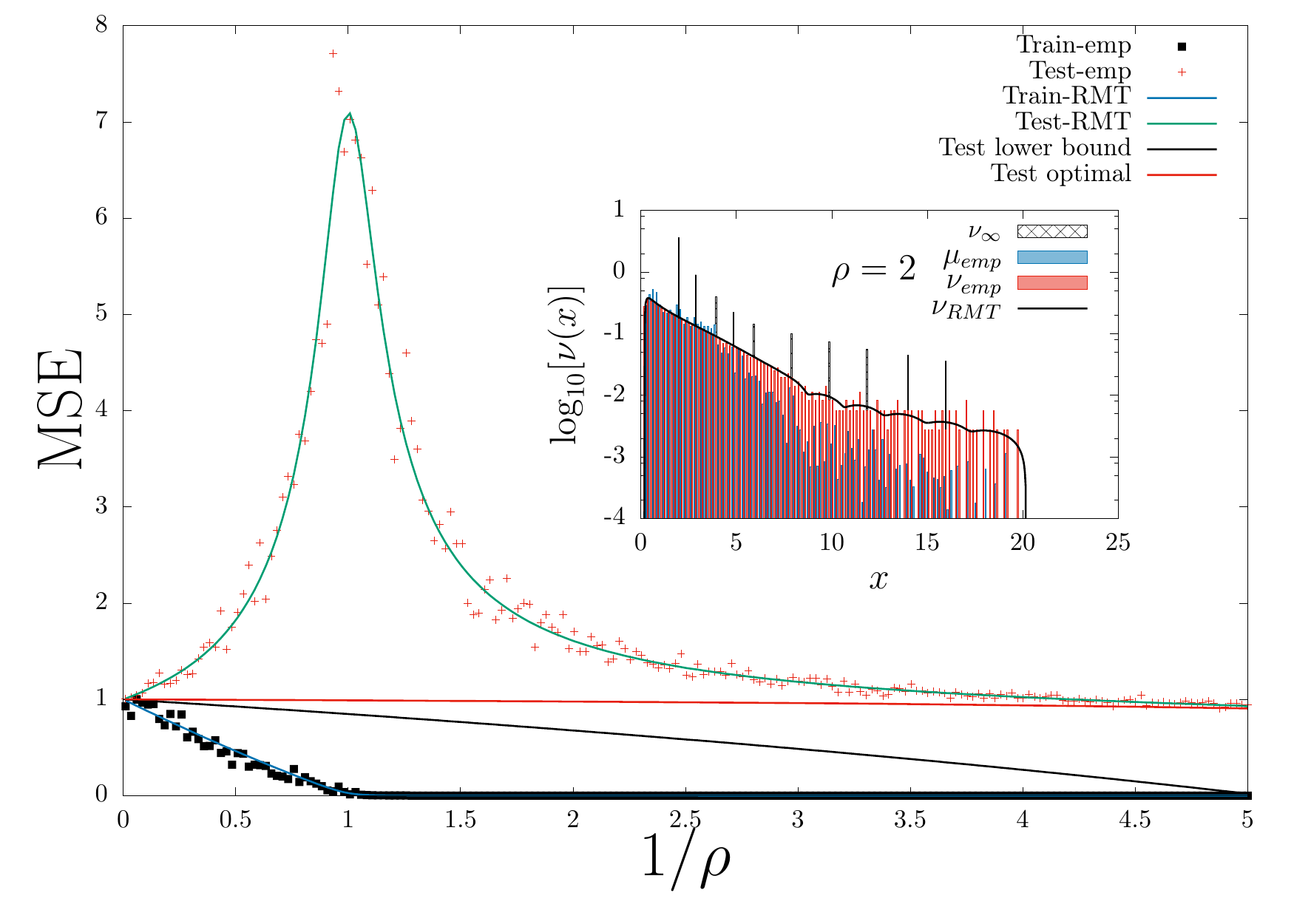}}
\centerline{\includegraphics[width=0.7\textwidth]{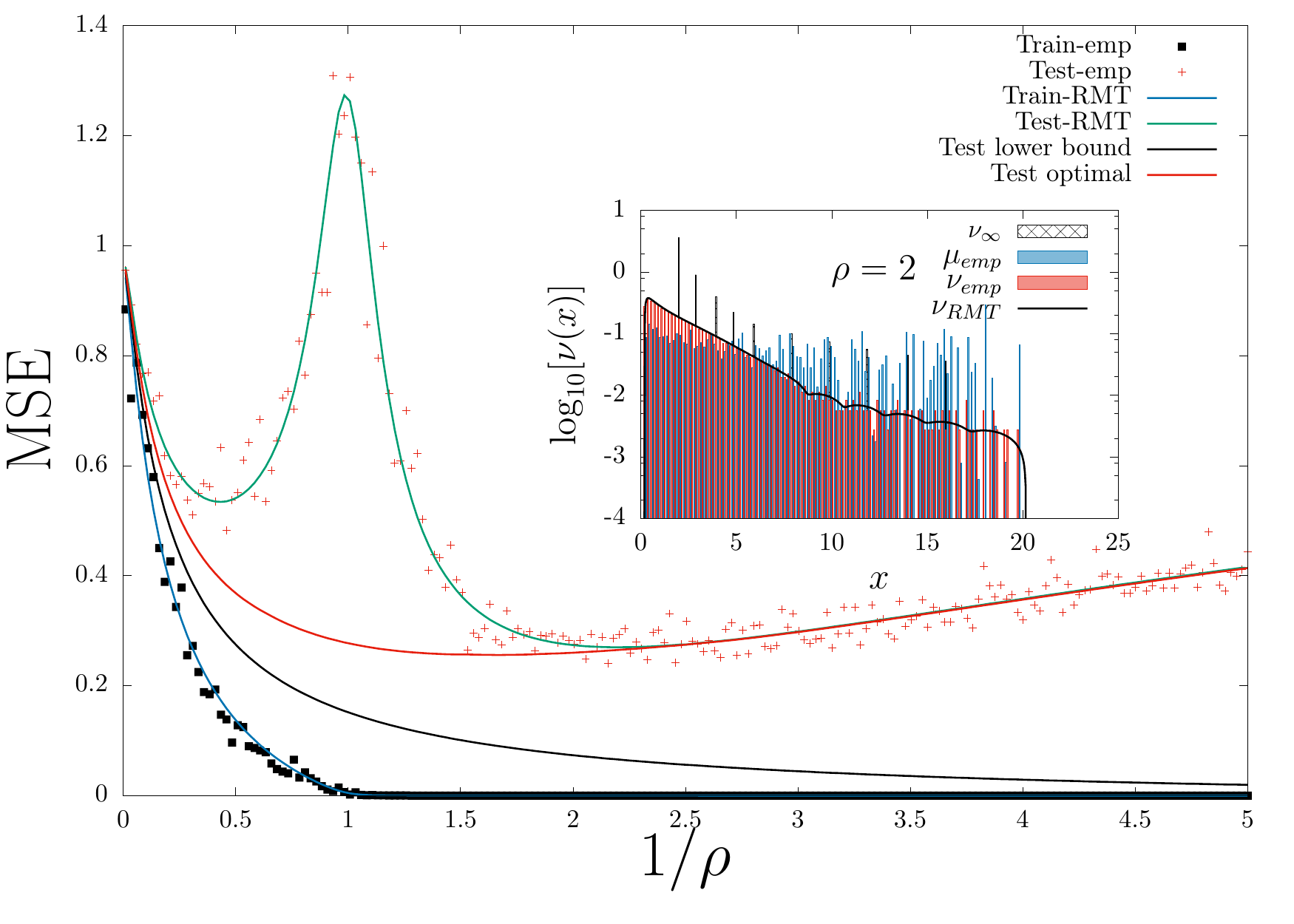}}
\caption{\label{fig:Train-test-spectral} Main plots: Test and train error as a function of $\rho^{-1}$ given by RMT (plain lines) and a single experiments with $N=200$,
  when the population matrix has one degenerate level (top panel) and ten degenerate levels,
  with wrong alignment (middle panel) and good alignment (bottom panel) of the signal with the population matrix. Inset: spectral densities of the population (in gray), empirical feature matrices (in red)
  and spectral power of the signal on the corresponding levels (blue).}
\end{figure}
We are now in position to simplify the asymptotic expression of the test error given in the preceding section. First we introduce the fraction $r$ of the signal that can be possibly recovered 
\begin{equation}\label{def:r}
r = \frac{\mu^\parallel}{\mu^\parallel+\mu^\perp+\sigma^2} = \frac{SNR}{1+SNR}
\end{equation}
where
\[
SNR \egaldef \frac{\mu^\parallel}{\mu^\perp+\sigma^2}
\]
is the effective signal to noise ratio, from the definition~(\ref{eq:sigma_eff}) of the effective noise variance. In order to have a meaningful comparison 
between various situation we normalize the errors by $\mu^\parallel+\mu^\perp+\sigma^2$, or equivalently assume  that the observation $y$ in~(\ref{eq:teacher}) is normalized s.t. 
\begin{equation}\label{eq:signalnorm}
\mu^\parallel+\mu^\perp+\sigma^2 = 1.
\end{equation}
We have
\begin{align*}
  \frac{\partial}{\partial \alpha} g(x,\rho,\alpha) &= - \frac{1}{\rho}\frac{(\rho+g[\Lambda]-1)^2}{\rho-\gd[\Lambda]}\frac{x}{\bigl(1+\Lambda x\bigr)^2}\\[0.2cm]
  \frac{\partial}{\partial \alpha} \bigl[\alpha g(\rho,\alpha)\bigr]
  &= g[\Lambda] - \frac{\rho-1+g[\Lambda]}{\rho-\gd[\Lambda]}\bigl[1-g[\Lambda]-\gd[\Lambda]\bigr],
\end{align*}
with the shorthand notation $\Lambda = \Lambda(\rho,\alpha)$.
Combining this with~(\ref{eq:FP}) in~(\ref{eq:Rasymp},\ref{eq:Etrainasymp}) and arranging the various terms leads us to our final expression for the (normalized) test error
\begin{equation}
  E_{\rm test}[\rho,r,\Lambda] = \frac{\rho}{\rho-\gd[\Lambda]}\bigl(1-r+r\mu_2[\Lambda]\bigr)\qquad \in[1-r,1] \label{eq:Etestasymp}
\end{equation}
expressed now as a function of $\rho$, $r$ and $\Lambda$. The contributions corresponding respectively to the bias and variance are then identified as
\begin{align}
  \mathcal{B}[\rho,r,\Lambda] &= \frac{\rho}{\rho-\gd[\Lambda]}r\mu_2[\Lambda],\label{eq:bias} \\[0.2cm]
  \mathcal{V}[\rho,r,\Lambda] &= \frac{\rho}{\rho-\gd[\Lambda]}(1-r).\label{eq:variance}
\end{align}
The expressions which appeared recently in statistics~\cite{dobriban2018high,hastie2022surprises} or ML literature~\cite{wu2020optimal,richards2021asymptotics},
are either less general or/and maybe more obscure to us. At least we could check that ours coincide with that of~\cite{dobriban2018high} when restricting the signal's distribution
to an isotropic one. Then looking at~(\ref{eq:Etestasymp}) we immediately see two counter effects from the regularization to be at work. One comes from the features and the other one from the signal
and are simply expressed by the spectral coefficients~(\ref{def:QLambda}) and~(\ref{def:mu_k}) with $k=2$. When the coupling is increased,
\[
\gd'[\Lambda] = 2\bigl(g[\Lambda]+g_3[\Lambda]-2g_2[\Lambda]\bigr) = 2\int dx\nu_\infty(x)\frac{\Lambda x^2}{(1+\Lambda x)^3} > 0,
\]
so $\gd$  increases, while $\mu_2$ obviously decreases, inducing the two counter effects on $E_{\rm test}$; the first contributes to some increase, the latter to a decrease,
and the optimal tuning of the penalty corresponds to the best trade-off between these two.
We also immediately see, as expected,  that $E_{\rm test}$ is
lower bounded by $1-r$ ($\mu_2 = \gd = 0$). In order to get as close as possible from this lower bound, as stated in~\cite{wu2020optimal},
a favorable situation corresponds to have a good alignment of the population matrix with the signal,
meaning that the signal decomposes preferentially on the stronger modes of the population matrix. Equation~(\ref{eq:Etestasymp}) allows us to have more precise indications on the
necessary ingredients for the ideal situation concerning generalization. Ideally we want the signal to decompose mainly on a small fraction $\nu_s$ of strong modes of the population
matrix, with $\nu_s\ll\rho$  separated by a large gap from  the other modes. This situation offers then a good choice for the ridge penalty which is to select $\Lambda^{-1}$ somewhere in the middle of the gap.
Doing that insures on one side that $\Lambda x \ll 1$ on the weak modes and on the other side that $\Lambda x\gg 1$ on the strong modes; in turn
this insures that $\gd\sim\nu_s \ll \rho$ and $\mu_2\ll 1$. This mechanism is illustrated on Figure~\ref{fig:Train-test-spectral} where we also see that up to fluctuations due mainly to the
finite number $N_s=200$ of train samples  (the number of test samples $>10^4$ is chosen sufficiently large  to have no influence on the variance of the results), the test and train error
are already properly estimated from RMT results given above. An example of a double descent is seen on the bottom panel, but as shown from the curve corresponding to optimal penalty this
is an artifact of a wrongly tuned regularization, which is kept constant while varying $\rho$.
In the next sections of the paper we will exploit this formulation which will appear very convenient when discussing the feature learning process.

\subsection{Practical aspects from empirical estimations}\label{sec:practical}
\begin{figure}
\centerline{\includegraphics[width=0.7\textwidth]{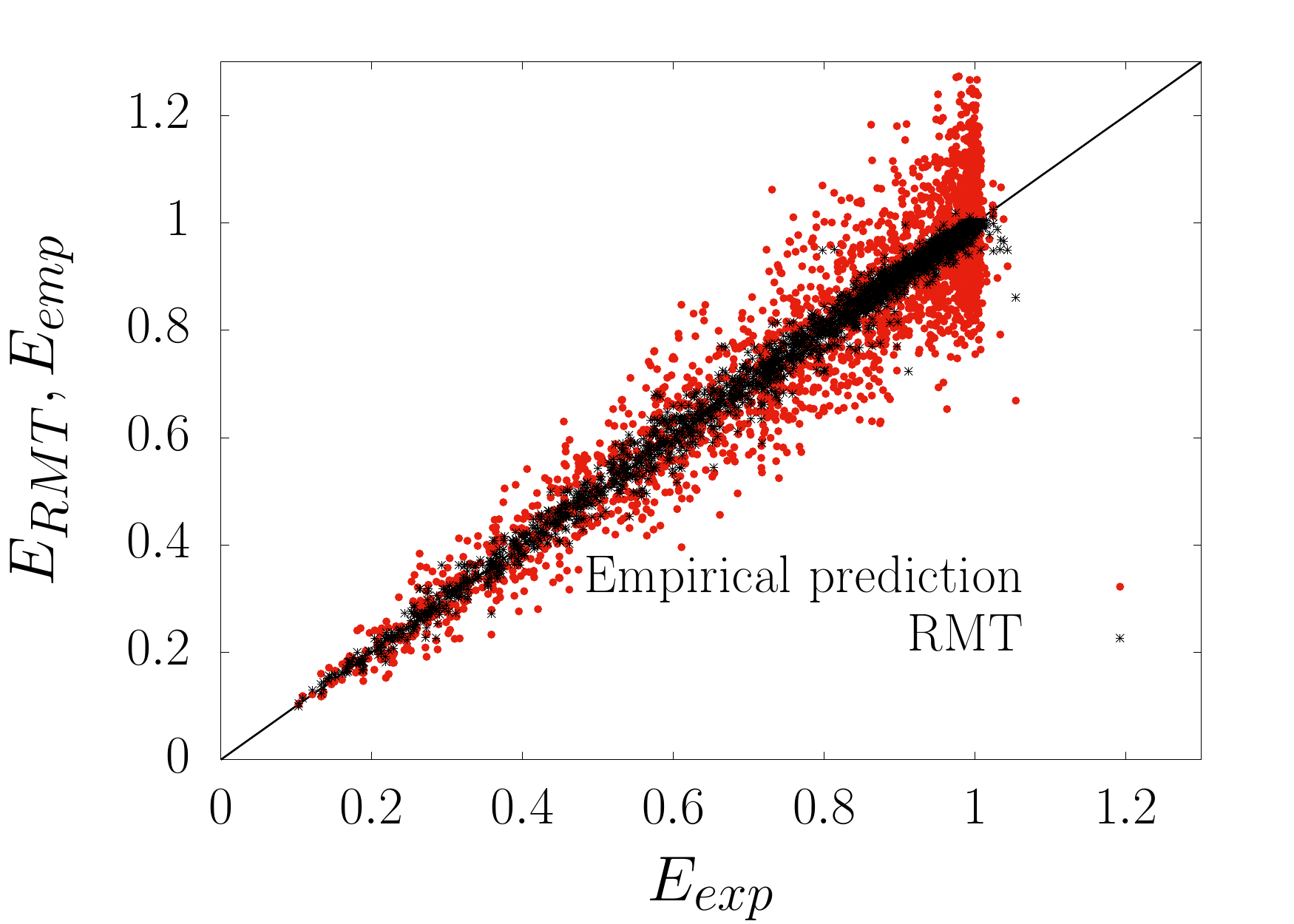}}
\caption{\label{fig:perf-diag-scatter} Scatter plot of the test error $E_{\rm exp}$ measured on experiments  done with a two $2$-level population matrix model with eigenvalues $\{1,50\}$,
  versus the expected results~(\ref{eq:Etestasymp}) from RMT  and the empirical prediction~(\ref{eq:E_emp}) obtained from the training.
  The experimental point are obtained with a train set of size $N_s=200$ and test set of size $10^4$. By convenience, the default coupling is
  set to $\alpha=\alpha^\star$ delivered by the RMT, $\rho$ is varied $\in\{0.5,1.0,2.0\}$, $\nu\in\{0.01,0.99\}$ and $\mu\in[0,1]$. 
}
\end{figure}

Since~(\ref{eq:Etestasymp}) is written as function of $\rho$, $r$ and $\Lambda$, finding the optimal ridge penalty for the test error is equivalent to
optimize $E_{\rm test}$ w.r.t. $\Lambda$ at fixed $\rho$ and $r$. Doing this yields the following condition on $\Lambda$:
\begin{equation}\label{eq:Lambdaopteq}
E_{\rm test}[\rho,r,\Lambda]\int dx \nu_\infty(x)\frac{\Lambda^2 x^2}{\bigl[1+\Lambda x\bigr]^3}-\frac{\rho r}{\mu^\parallel}\int dx\mu^\parallel(x)\frac{\Lambda x}{\bigl[1+\Lambda x\bigr]^3} = 0.
\end{equation}
It is not clear yet how to solve this equation in practice. 
Just notice here that the previous condition gives us a way to parameterize the optimal $E^\star_{\rm test}$ in terms of the optimal value $\Lambda^\star$ of the effective coupling as
\[
E_{\rm test}^\star[\rho,r] = \rho r\ \frac{\mu_2[\Lambda^\star]-\mu_3[\Lambda^\star]}{g[\Lambda^\star]+g_3[\Lambda^\star]-2g_2[\Lambda^\star]}
\]
written in terms of the spectral coefficients defined in~(\ref{def:g_k},\ref{def:mu_k}) when $\Lambda^\star$ is finite.
Yet there are considerations of more practical interest  that can be
deduced from the asymptotic theory in complement of some others recently proposed  in~\cite{Liu2020Ridge} for instance.
First from~(\ref{eq:train-test},\ref{def:Lambdaeff}) and (\ref{eq:FP}) we see that
the estimation 
\[
g[\Lambda] \approx \hat g[\alpha] \egaldef \frac{1}{N_f}\Tr\Bigl[\frac{1}{\I_\F+\alpha C\n}\Bigr],
\]
directly made on the training data gives us an estimation of the effective coupling
\[
\Lambda \approx \hat\Lambda = \frac{\alpha}{\rho}\bigl(\rho-1+\hat g[\alpha]\bigr)
\]
and hence an estimation of the train-test error ratio, so that finally from the train
error we can get a prediction of the test error without resorting to cross-validation:
\begin{equation}\label{eq:E_emp}
\hat E_{\rm test} = \frac{\rho^2}{\bigl(\rho-1+\hat g[\alpha]\bigr)^2} E_{\rm train}.
\end{equation}
Again, a similar proposal is  made in~\cite{malzahnOpper2005} to exploit asymptotic relations between test and train error in the context of Gaussian processes.
As seen on the plot of Figure~\ref{fig:perf-diag-scatter}, a good correlation between this estimation and the actual test error is observed,
with a variance increasing with errors.

Nevertheless it is interesting to realize that
other statistics concerning the hidden prior, namely the population matrix, are accessible via other estimators in this asymptotic regime.
For instance the coefficient $g_2$ can be estimated as well. 
Let
\[
\hat g_2(\alpha) \egaldef \frac{1}{N_f}\Tr\Bigl[\frac{1}{\bigl(\I_\F+\alpha C\n\bigr)^2}\Bigr].
\]
On the one hand we have
\[
\frac{d}{d\alpha}\hat g(\alpha) = \frac{1}{\alpha}\bigl[\hat g_2(\alpha)-\hat g(\alpha)\bigr].
\]
On the other hand
\[
\frac{d}{d\alpha} g[\Lambda] = \frac{d\Lambda}{d\alpha}g'[\Lambda] = \frac{1}{\Lambda}\frac{d\Lambda}{d\alpha}\bigl(g_2[\Lambda]-g[\Lambda]\bigr). 
\]
Identifying the two derivatives and making use of~(\ref{eq:FP},\ref{eq:dLambda}), we can rearrange things to obtain the following estimator for $g_2$:
\[
g_2[\Lambda] \approx  \hat g + \frac{\hat g_2 - \hat g}{1+\frac{\hat g_2 - \hat g}{\rho-1+\hat g}}.
\]
In principle we should be able to carry on this program for $g_k, k>2$ which eventually is equivalent to trying to solve the inverse problem of the population 
matrix given the sample covariance matrix~\cite{bun2017cleaning}. Typically, if $C$ has $K$ distinct levels, we have to go until $k=K$ to be able to resolve all its levels.
For instance, the $2$-level problem considered in the next Section~\ref{sec:Specialcases} has two unknowns regarding the population matrix, namely $\nu$ the degeneracy of
the level associated to the strong modes and $\Delta$ the gap between the two levels. Obviously the knowledge of $g$ and $g_2$ is enough to find these two
unknowns. While this is an interesting direction relevant to practical ML applications,
it is out of the scope of the present paper and we leave it aside for future works.

Concerning the signal itself, it is also possible to extract information. Similarly we would like to be able to estimate the coefficients $\mu_k$, at least for
small $k$, but also $\mu^{\parallel,\perp}$.
First the mean-squared norm of the observed signal is given empirically by
\begin{equation}\label{eq:ynorm}
\mu^\parallel+\sigma_{\rm eff}^2 \approx \frac{1}{N}\sum_{s=1}^{N_s} {y^{(s)}}^2.
\end{equation}
Then the train error normalized by this quantity is already a source of information since its expression in terms of $\mu_k$ coefficients actually read
\begin{equation}\label{eq:Etrainasymp2}
E_{\rm train} = \frac{\rho\Lambda^2}{\alpha^2\bigl(\rho-\gd[\Lambda]\bigr)}\bigl[r\mu_2+(1-r)\bigr] 
\end{equation}
where here $\Lambda$ and $\gd[\Lambda] = 1-2g[\Lambda]+g_2[\Lambda]$ can be estimated empirically from the preceding remarks, hence the unknown are $r$ and $\mu_2$.
Another interesting quantity is the squared norm $\Vert\w\Vert^2$ of the regression vector~(\ref{eq:ridge_sol}),
\[
\w^t\w = \f^t F^+\bigl(\I-G\n\bigr)^2F^{+t}\f+\frac{\alpha\sigma_{\rm eff}^2}{N}\Tr\Bigl[G\n\bigl(\I-G\n\bigr)\Bigr],
\]
which when normalized by~(\ref{eq:ynorm}) is given asymptotically by
\[
\w^t\w = r\Lambda\Bigl[\mu_1-\frac{\rho\Lambda}{\alpha\bigl(\rho-\gd[\Lambda]\bigr)}\mu_2\Bigr]+(1-r)\Lambda\Bigl[1-\frac{\rho\Lambda}{\alpha\bigl(\rho-\gd[\Lambda]\bigr)}\Bigr].
\]
Combining these two expressions yields the asymptotic expression of the normalized loss:
\begin{equation}
  \LL = \frac{1}{2}\bigl(\w^t\w+\alpha E_{\rm train}\bigr)
  = \frac{\Lambda}{2}\bigl[1+r(\mu_1-1)\bigr],\label{eq:Lossasymp}
\end{equation}
which will be used to determine the dynamics in Section~\ref{sec:learningdynamics}.
Hence combining the empirical measurement of $\w^t\w$ and of $E_{\rm train}$
allows us to disentangle $\mu_1$ from $\mu_2$ when $r=1$, i.e. in absence of effective noise. In this case, derivation w.r.t. $\alpha$ would let us obtain empirical estimation
of higher order $\mu_k,k>1$ coefficients. When noise is present or whenever the model is misspecified a possible way to proceed, but not explored here, would consist in probing
the system by adding some noise and extract empirically the rate of dependence of $\LL$ to that noise, namely $\Lambda(1-\mu_1)/2$ as seen from~(\ref{eq:Lossasymp}).
\begin{figure}
\centerline{\includegraphics[width=0.7\textwidth]{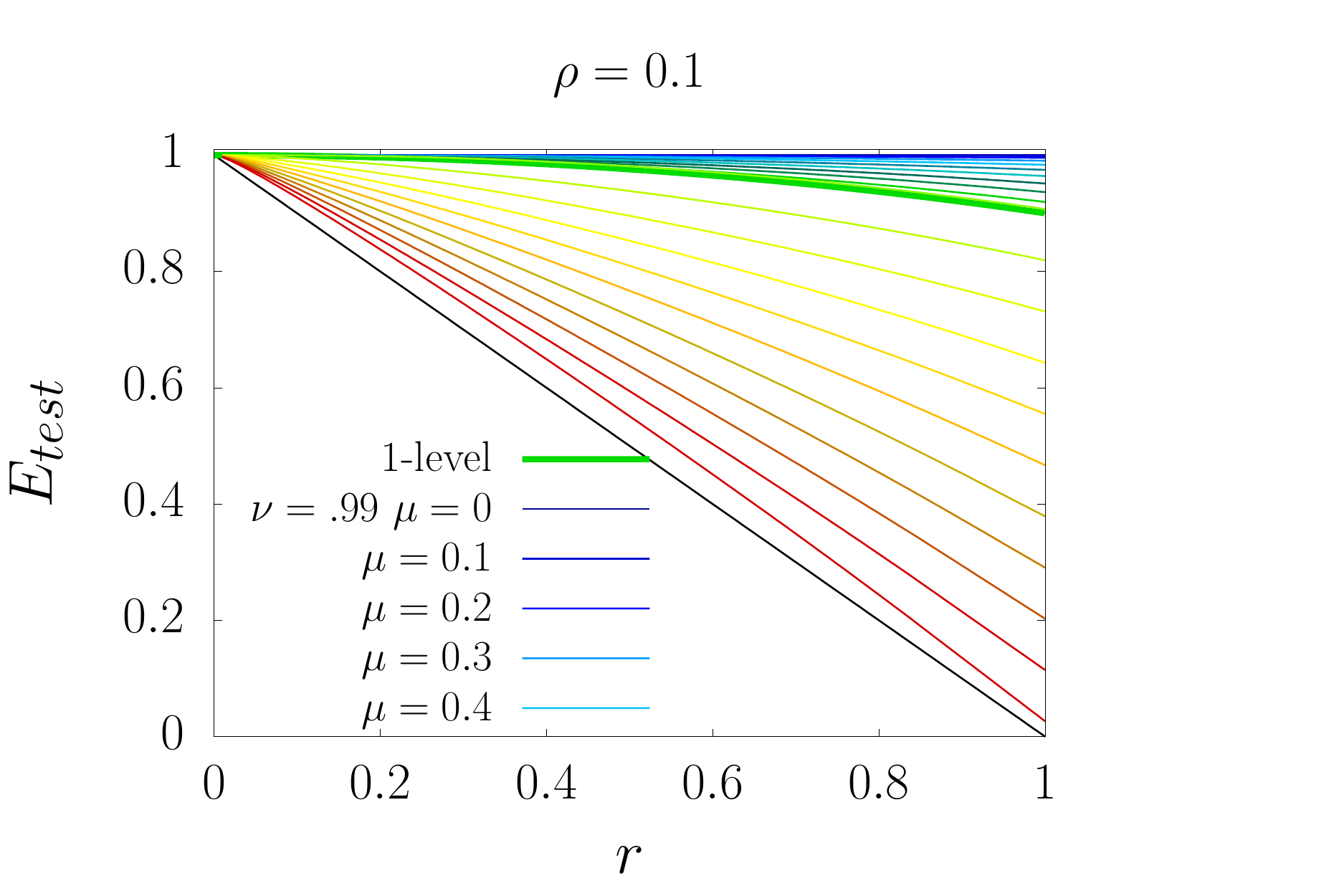}\hspace{-2cm}
\includegraphics[width=0.7\textwidth]{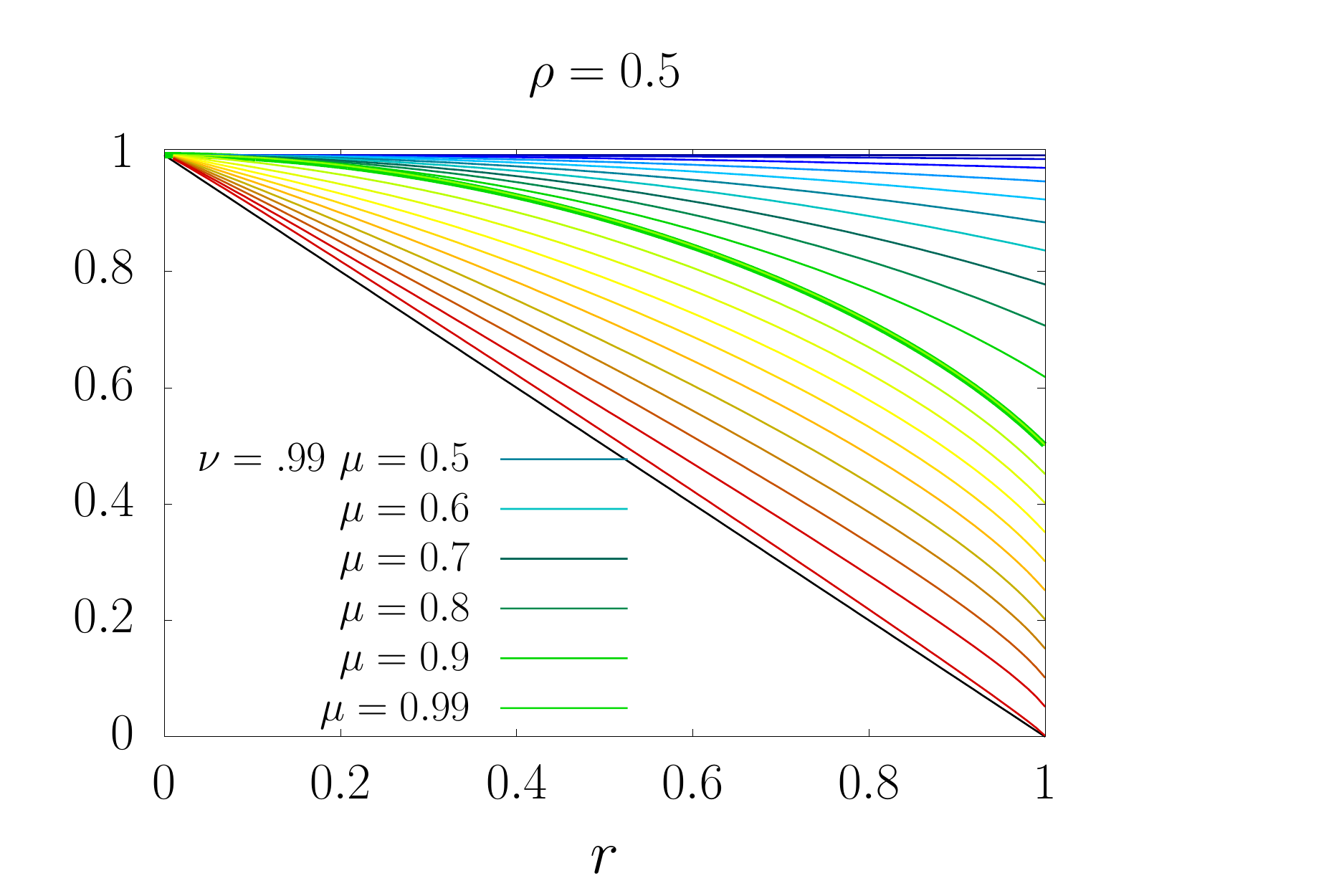}}
\centerline{\includegraphics[width=0.7\textwidth]{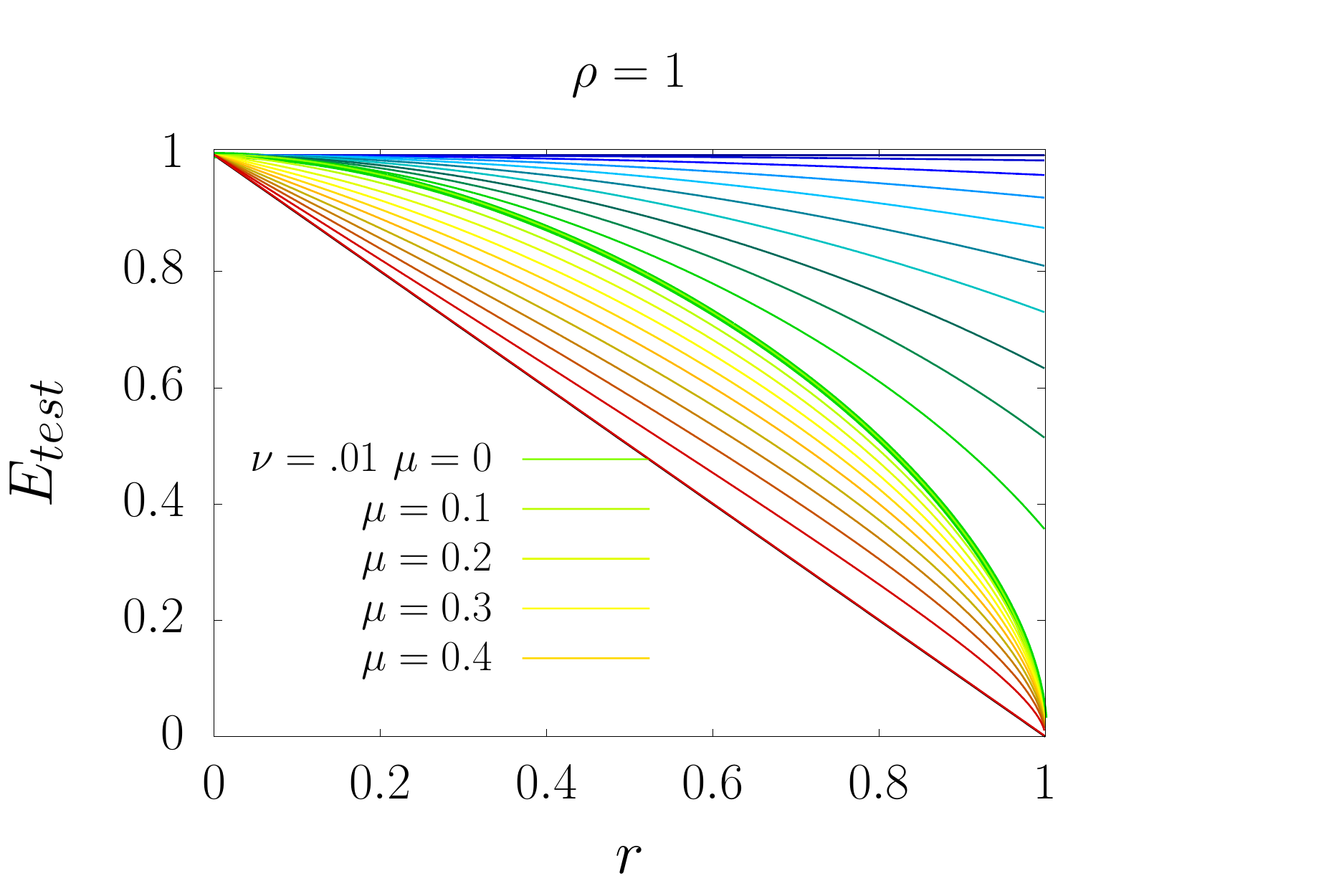}\hspace{-2cm}
\includegraphics[width=0.7\textwidth]{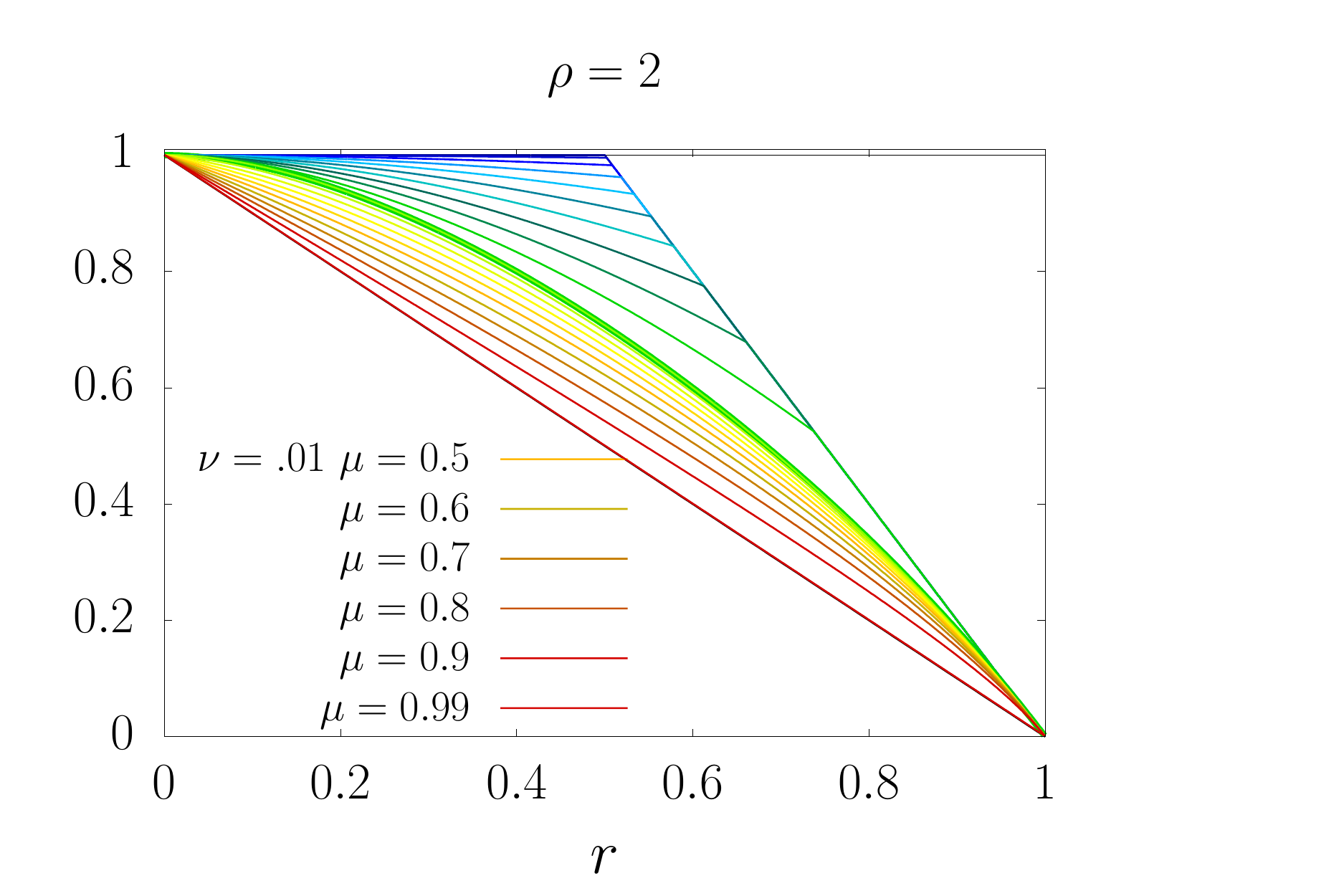}}
\caption{\label{fig:perf-diag} The optimal performance in terms of $E_{\rm test}$ of the $2$-level model as a function of $r$, depending on the fraction $\nu$ of strong modes and
  on the fraction $\mu$ of longitudinal spectral power of the signal on these modes. The performance of the $1$-level model  defines
  the frontier between the aligned and anti-aligned domains, represented here respectively by the $\nu=0.01$ and $\nu=0.99$ set of curves. 
}
\end{figure}
\subsection{Special cases}\label{sec:Specialcases}
To illustrate the preceding statements let us consider  some special cases.
\paragraph{The one-level case:} it 
corresponds to the fully degenerate case, when the population matrix $C=\I$, i.e. when 
\[
\nu_\infty(x) = \delta(x-1).
\]
This case corresponds for instance to the functional case where the features are normalized Fourier modes along with a uniform spatial distribution of input sample points.
The relative longitudinal spectral density of the signal is then necessary of the form
\[
\mu^\parallel(x) = \mu^\parallel\delta(x-1),
\]
so that we have the form $g_k = \mu_k = y^k$ with $y\in[\max(0,1-\rho),1]$
yielding
\begin{equation}\label{eq:Etestsingle}
  E_{\rm test} =  \frac{\rho}{\rho-(1-y)^2}\bigl[1+r(y^2-1)\bigr].
\end{equation}
The optimization can be done here directly on $y$ (which is a monotonic function of $\Lambda$) and yields
\begin{equation}\label{eq:gstar}
  y^\star = \frac{1}{2r}\Bigl[ 2r+\sqrt{(1+\rho r)^2-4\rho r^2}-1-\rho r\Bigr],
\end{equation}
giving the optimal  test error as a function of $r$ in the form
\[
E_{\rm test}^\star[\rho,r] = \rho r \frac{2r+\sqrt{(1+\rho r)^2-4\rho r^2}-1-\rho r}{1+\rho r-\sqrt{(1+\rho r)^2-4\rho r^2}},
\]
plotted on Figure~\ref{fig:perf-diag}.

\paragraph{Two-level case:} it corresponds now to the case (also considered in~\cite{richards2021asymptotics}) were the spectrum of the population matrix has two levels representing weak and strong modes.
Note that the problem is invariant when multiplying the ridge penalty and the population matrix by a common factor, so  
in great generality  for this case we can consider a spectral density of the form
\[
\nu_\infty(x) = (1-\nu)\delta(x-1)+ \nu\delta(x-c),
\]
with $c>1$. $\nu\in[0,1]$ represents the fraction of strong modes.  

Then the relative spectral density of the signal  takes the form
\[
\mu^\parallel(x) = \mu^\parallel\bigl[(1-\mu)\delta(x-1)+\mu\delta(x-c)\bigr],
\]
where $\mu\in[0,1]$ represents the weight of the signal on the strong modes.
Given this we obtain a test error given by~(\ref{eq:Etestasymp}) with now
\begin{align*}
  \mu_2[\Lambda] &= \frac{1-\mu}{\bigl[1+\Lambda\bigr]^2}+\frac{\mu}{\bigl[1+\Lambda c\bigr]^2}, \\[0.2cm]
  \gd[\Lambda] &= (1-\nu)\frac{\Lambda^2}{\bigl[1+\Lambda\bigr]^2}+\nu \frac{\Lambda^2 c^2}{\bigl[1+\Lambda c\bigr]^2}.
\end{align*}
Looking for the optimal $\Lambda^\star$ amounts to solve~(\ref{eq:Lambdaopteq})  as a complicated polynomial equation.  
To simplify let us consider the case of a large gap, i.e. $c \gg 1$. In that case we can distinguish two distinct asymptotic regimes when $c\to\infty$.
First from  the discussion of the end of Section~\ref{sec:LambdaEff} we may expect
$\Lambda^{-1}$ to scale like $c$, i.e. $\Lambda c = {\mathcal O}(1)$. Letting
\[
y \egaldef \frac{1}{1+\Lambda c}\in[0,1],
\]
we obtain
\begin{align*}
  \mu_2(\eta) &= 1-\mu+\mu y^2 +{\mathcal O}\Bigl(\frac{1}{c}\Bigr),  \\[0.2cm]
  \gd(\eta) &= \nu (1-y)^2 +  {\mathcal O}\Bigl(\frac{1}{c^2}\Bigr),
\end{align*}
and the test error then reads
\[
E_{\rm test} = \frac{\rho}{\rho-\nu (1-y)^2}\bigl[1+r\mu(y^2-1)\bigr], 
\]
which derivative w.r.t. $y$ vanishes at
\[
y^\star = \frac{1}{2a}\Bigl[\sqrt{\bigl(1+a\eta\bigr)^2+4a(1-a)}-1-a\eta\Bigr],
\]
with $a\egaldef r\mu$ and $\eta\egaldef \rho/\nu-2$.
The second asymptotic regime corresponds to the situation where $\Lambda c\to\infty$, in which case $\Lambda$ is not necessarily vanishing.
Letting now
\[
y \egaldef \frac{1}{1+\Lambda}\in[0,1],
\]
we obtain
\begin{align*}
  \mu_2(\eta) &= (1-\mu) y^2 +{\mathcal O}\Bigl(\frac{1}{c}\Bigr),  \\[0.2cm]
  \gd(\eta) &= (1-\nu) (1-y)^2 + \nu + {\mathcal O}\Bigl(\frac{1}{c}\Bigr),
\end{align*}
and the test error then reads
\[
E_{\rm test} = \frac{\rho}{\rho-\nu-(1-\nu)(1-y)^2}\bigl[1-r+r(1-\mu)y^2\bigr].
\]
This expression actually coincides with the previous one up to a constant factor when doing the change $r\to r/(1-r\mu)$, $\rho\to \rho-\nu$, $\nu\to 1-\nu$ and $\mu\to 1-\mu$,
leading to the same $y^\star$ in terms of the modified parameters $a$ and $\eta$. Note that in this regime for $\rho<1$, the constraint $g>1-\rho$ becomes $y> \frac{1-\rho}{1-\nu}$,
hence for $\nu>\rho$, $y_{\rm min}>1$ so this asymptotic regime is irrelevant and only the first one needs to be considered. Otherwise, given the two minimal values
of $E_{\rm test}$ obtained, the lowest is selected. The result of this is shown on Figure~\ref{fig:perf-diag}, where we see that optimal performance 
is obtained with the perfect alignement when the signal is concentrated on very few strong modes, i.e. $\nu\to 0$ and $\mu\to 1$. The $1$-level case
appears to separate the good from bad alignement settings in these plots.

\section{Learning dynamics of the population matrix}\label{sec:learningdynamics}
\begin{figure}
\centerline{\resizebox{0.8\textwidth}{!}{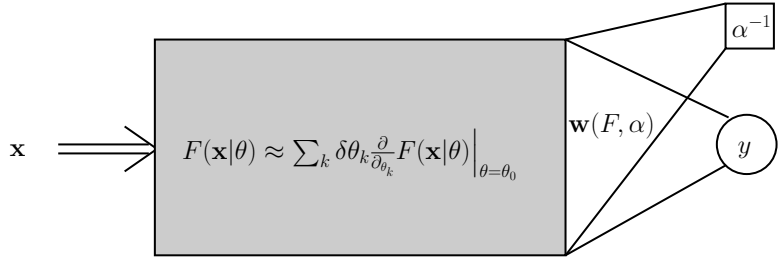}}
\caption{\label{fig:student} Schematic plot of the regression model considered here. The weight vector $\w(F,\alpha)$ of the last layer solves a ridge regression with penalty $\alpha^{-1}$
  depending on the vector of features functions $F(\x\vert\theta)$ defined by some NN with parameters $\theta$ represented by the grey box.
  In the ``semi-lazy'' regime $F(\x\vert\theta)$ is taken close to some reference parameter $\theta_0$, with $\delta\theta=\theta-\theta_0$.
}
\end{figure}

Now we are in position to analyze the learning dynamics in a more general setting corresponding for instance to neural networks (NN) regressions.
Dynamics of neural networks have been considered in several works, in particular in~\cite{Saxe-Ganguli_2013} is characterized the (non-linear) dynamics of deep
linear neural networks, or in~\cite{saxe2022neural} (and ref. herein) for non-linear ones.
Here we study something close but different in spirit, by considering the architecture sketched on Figure~\ref{fig:student},
where at the first stage, features functions are provided by some arbitrary NN which don't need to be specified for the moment, while the last layer
is linear and combine these features to solve a ridge regression. As a result, the weights of the last layer are given in closed form by~(\ref{eq:ridge_sol})
with explicit dependency on the feature matrix. Our goal is to study whether the learning process leads to a better alignment of the features with the signal as the one chosen
randomly at initialization.
\subsection{Microsopic equations of learning}\label{sec:micro}
To this end let us reconsider the misspecified regression problem~(\ref{eq:mispecified}), with a feature matrix being now a function $F(\x\vert\theta)$
of the input data $\x$ and of some vector parameter $\theta$ of a NN, while the vector $\w$ constitutes the weight vector of the last layer
\[
f(\x\vert \w,\theta) = \w^t F(\x\vert\theta).
\]
We keep the $L_2$ penalty on the vector $\w$ in the loss function
which is there to mimic some form of implicit regularization of the whole architecture:
\begin{equation}\label{eq:loss2}
\LL(\w,\theta) = \frac{1}{2}\w^t\w +\frac{\alpha}{2N_s}\sum_{s=1}^{N_s}\bigl[y^{(s)}-\w^tF(\x^{(s)}\vert\theta)\bigr]^2.
\end{equation}
In order to study the dynamics of the features under the gradient descent we consider a continuous limit $\theta_t$ of the learning process indexed by $t$
and assume that $\w= \w(\theta_t)$ moves  ``adiabatically'' along the gradient descent of $\theta$,
meaning that we use~(\ref{eq:ridge_sol})
\[
\w(\theta) = \alpha G\n(\theta) Z\n(\theta),
\]
with now $G\n(\theta) = \bigl[\I_\F+\alpha C\n(\theta)\bigr]^{-1}$ and
\begin{align}
  C\n(\theta) &\egaldef \frac{1}{N}\sum_{s=1}^N F(\x^{(s)}\vert\theta) F^t(\x^{(s)}\vert\theta),\label{eq:CN}\\[0.2cm]
  Z\n(\theta) &\egaldef   \frac{1}{N}\sum_{s=1}^N F(\x^{(s)}\vert\theta) y^{(s)}.\label{eq:ZN}
\end{align}
The dynamics~(\ref{eq:thetadot}) is modified as ($\gamma=1$)
\[
\dot\theta_t = -\nabla_\theta\LL\bigl[\w(\theta_t),\theta_t\bigr].
\]
Next we write the singular value decomposition of the feature matrix as
\[
F(\cdot\vert \theta) = \sum_{a=1}^{N_f}\sqrt{c_a(\theta)} \bu_a(\theta)v_a(\cdot\vert\theta),
\]
where the $v_a$ which  depends on $\theta$, are elements of a functional basis on the input space $\R^d$. This is the counterpart of~(\ref{eq:Fsvd})
written for a functional space, which dimension $D$ grows potentially exponentially with the dimension $d$ of the base  space.  
We assume them to be orthonormal with respect to the scalar product relative to the prior density $\rho_\infty(\x)$ of the data:
\[
\int d^d\x \rho_\infty (\x) v_a(\x) v_b(\x) = \delta_{ab}.
\]
The hidden function $f(\x)$ may be considered indifferently  through its vector representation $\f = \{f_a\}$ on this vector space,
with 
\[
f(\x) = \sum_{a=1}^D f_a v_a(\x)\qquad\text{and}\qquad
f_a = \int d^d\x \rho_\infty (\x) v_a(\x) f(\x).
\]
The population matrix obtained by taking $N\to\infty$ in~(\ref{eq:Cn}) reads as before
\[
C(\theta) = \sum_{a=1}^{N_f} c_a(\theta) \bu_a(\theta)\bu_a^t(\theta),
\]
thanks to the hypothesis that $\rho_\infty$ is uniform, which is the counterpart in continuous space to the one considered in~(\ref{eq:popmatrix}). 
This stated we can now express the evolution of the feature matrix by its derivative w.r.t. $t$ on the basis $\{\bu_a\}$ and $\{v_a(\x)\}$. In order to lighten the notations
we will use $\bv_a$ instead of $v_a(\x)$ to denote the right functional basis and hide any explicit dependency over $\x$. 
We follows the same procedure as what was done previously to model the learning dynamics of restricted Boltzmann machines, approximately for the binary-binary in~\cite{DeFiFu_2017,DeFiFu_2018}
or exactly for the Gaussian-spherical in~\cite{DeFu2020}. Infinitesimal variations of the feature matrix are given by
\begin{equation}
  dF = \sum_{a=1}^{N_f} \Bigl[\frac{dc_a}{2\sqrt{c_a}}\bu_a\bv_a^t+ c_a\sum_{b=1}^{N_f}\Bigl( d\Omega_{ab}^{(u)}\bu_b\bv_a^t +d\Omega_{ab}^{(v)}\bu_a\bv_b^t\Bigr)\Bigr],\label{eq:dF}
\end{equation}
where $d\Omega^{(u,v)}$ are skew-symmetric operators representing infinitesimal rotation generators of the $\bu_a$ and $\bv_a$ vectors,
\begin{align*}
  d\Omega_{ab}^{(u)} \egaldef (d\bu_a^t) \bu_b =  -d\Omega_{ba}^{(u)},\\[0.2cm]  
  d\Omega_{ab}^{(v)} \egaldef (d\bv_a^t) \bv_b =  -d\Omega_{ba}^{(v)}.  
\end{align*}
At this point let us remark that the problem enjoys a gauge symmetry, namely the loss function is invariant under the combined transformation 
\[
(\w,F) \longrightarrow (R^t\w,R F),
\]
where $R = e^{\Omega^{(u)}}$ is any left rotation, i.e. on feature space, so that we may focus on the effect of scalar transformations and right rotations.
The scalar transformation corresponds to modifications of the $\{c_a\}$ i.e. modifying the relative strength between the features,
while the right rotations corresponds to changing the features themselves.
The relation~(\ref{eq:dF}) can be inverted, yielding in particular for right rotations (from now on we drop the $(v)$ index)
\begin{equation}\label{eq:dOmega}
d\Omega_{ab} = \frac{1}{\sqrt{c_a}+\sqrt{c_b}}\Bigl(\bu_a^t dF \bv_b\Bigr)^A + \frac{1}{\sqrt{c_a}-\sqrt{c_b}}\Bigl(\bu_a^t dF \bv_b\Bigr)^S,
\end{equation}
where $A,S$ denote respectively the anti-symmetric and symmetric parts w.r.t. $a,b$. In turn we can express infinitesimal transformations of the loss function under
such deformations. Using its asymptotic form~(\ref{eq:Lossasymp}) obtained in Section~\ref{sec:practical} written at finite $N_f$, i.e. with
\[
\Lambda = \frac{\alpha}{\rho}\Bigl(\rho-1+\frac{1}{N_f}\sum_{a=1}^{N_f} \frac{1}{1+\Lambda c_a}\Bigr)
\]
from~(\ref{eq:FP}), and assuming the normalization~(\ref{eq:signalnorm}) of the signal we have 
\[
r = \sum_{a=1}^{N_f} f_a^2 \qquad\text{and}\qquad r\mu_1 = \sum_{a=1}^{N_f} \frac{f_a^2}{1+\Lambda c_a}.
\]
Since the signal $\f$ is fixed  the effect of rotating the right (functional) bases induces variations of its components given by 
\[
df_a = \sum_{b=1}^{D}d\Omega_{ab}f_b.
\]
As a result we obtain
\begin{equation}\label{eq:dLambdaca}
  d\Lambda = -\frac{1}{N_f}\sum_{a=1}^{N_f} \frac{\Lambda^2}{(1+\Lambda c_a)^2}\frac{dc_a}{\rho-\gd},
\end{equation}
\[
  dr = \sum_{a,b=1}^{N_f,D} f_ad\Omega_{ab}f_b,
\]
and
\[
  d(r\mu_1) =  \sum_{a,b=1}^{N_f,D} \frac{f_ad\Omega_{ab}f_b}{1+\Lambda c_a} - \sum_{a=1}^{N_f}\frac{\Lambda}{(1+\Lambda c_a)^2}\bigl[f_a^2-\frac{r}{N_f}(\mu_1-\mu_2)\bigr]dc_a. 
\]
Using~(\ref{eq:dOmega}) we can now write the variations of the loss function in terms of $dF$ as ($dF_{aa} = dc_a/(2\sqrt{c_a}$)
\begin{equation}\label{eq:dLoss}
d\LL = -\sum_{a=1}^{N_f}\frac{\Lambda\sqrt{c_a}}{(1+\Lambda c_a)^2}\Bigl(f_a^2+\frac{{\mathcal B}}{N}\Bigr) dF_{aa} -
\frac{\Lambda^2}{2}\sum_{a,b=1}^{N_f,D}\frac{f_a f_b\sqrt{c_a}}{(1+\Lambda c_a)(1+\Lambda c_b)}dF_{ab},
\end{equation}
letting $c_b=0$ for $b>N_f$, ${\mathcal B}$ is the bias given by~(\ref{eq:bias}) and $N=\rho N_f$ the number of samples.
From these we can now obtain the equations expressing the learning dynamics asymptotically. In absence of constraints from the NN structure, to be discussed in the next Section,
the time variation of the feature matrix is simply given by the gradient descent of the loss function w.r.t. its coefficients 
\[
\frac{dF_{ab}}{dt} = -\frac{\partial\LL}{\partial F_{ab}},
\]
which can be read off from the previous equation.
First the scalar deformation gives us the dynamics of the eigenvalues of the population
matrix:
\[
\frac{d\sqrt{c_a}}{dt} = \frac{dF_{aa}}{dt}= -\frac{\partial\LL}{\partial F_{aa}}
\]
while the rotation of the eigenvectors are determined by the variation of off-diagonal coefficients of the feature matrix
\[
  \frac{d\Omega_{ab}}{dt} =  -\frac{1}{c_a-c_b}\Bigl(\sqrt{c_a}\ \frac{\partial\LL}{\partial F_{ab}}+
  \sqrt{c_b}\ \frac{\partial\LL}{\partial F_{ba}}\Bigr)\qquad\forall a\ne b.
\]
Again, $c_b=0$ by convention for transverse modes, i.e. $b> N_f$.
Note that the rotation between vectors belonging to the same degenerate eigenspaces is not defined and can be set to zero by convention.   
Consequently we get 
\begin{equation}
\frac{d\Omega_{ab}}{dt} = \Lambda^2 \frac{c_a+c_{b}}{c_a-c_{b}}\frac{f_af_{b}}{\bigl[1+\Lambda c_a\bigr]\bigl[1+\Lambda c_{b}\bigr]}.\label{eq:dotomega}
\end{equation}
The equations simplify by remarking that the transverse components are involved only through the macroscopic quantity~(\ref{def:r}). We have
\begin{equation}\label{eq:r}
r = \sum_{a=1}^{N_f} f_a^2 \qquad\text{and}\qquad \sum_{b=N_f+1}^D f_b^2 = 1-r-\sigma^2,
\end{equation}
thanks to the normalization~(\ref{eq:signalnorm}).
As a result we end up with the following dynamical system for $a=1,\ldots N_f$:
\begin{align}
  \dot c_a &=  \frac{2\Lambda^2 c_a}{\bigl[1+\Lambda c_a\bigr]^2}\Bigl(f_a^2+\frac{\mathcal{B}}{N}\Bigr),\label{eq:cat2} \\[0.2cm]
  \dot f_a &= \Lambda^2\Bigl[\frac{f_a}{1+\Lambda c_a}(1-r-\sigma^2)+\sum_{b=1,b\ne a}^{N_f}\frac{c_a+c_{b}}{c_a-c_{b}}\frac{f_af_{b}^2}{\bigl[1+\Lambda c_a\bigr]\bigl[1+\Lambda c_{b}\bigr]}\Bigr].
  \label{eq:fat2}
\end{align}
Given the definition~(\ref{eq:r},\ref{eq:bias}) of $r$ and the bias ${\mathcal B}$  
and the fact that $\Lambda$ is self-consistently determined through~(\ref{eq:FP}), all the macroscopic variables appearing in these equations can be expressed in terms of
spectral coefficients $g_k$ and $\mu_k$ given by~(\ref{def:g_k},\ref{def:mu_k}), which read  
\begin{equation}\label{eq:gk-muk}
g_k[\Lambda] = \frac{1}{N_f}\sum_{a=1}^{N_f}\frac{1}{(1+\Lambda c_a)^k},\qquad\text{and}\qquad \mu_k[\Lambda] = \sum_{a=1}^{N_f}\frac{f_a^2}{(1+\Lambda c_a)^k}
\end{equation}
when considered at finite size. Consequently the system~(\ref{eq:cat2},\ref{eq:fat2}) constitutes an autonomous dynamical system over the variables
$\{c_a,f_a;a=1,\ldots N_f\}$. It describes the evolution of the $2N_f$ microscopic degrees of freedom
of the regression problem.
As we see, the time derivatives of $c_a$ or $f_a^2$ are of order $1/N_f$,
while $c_a = \mathcal{O}(1)$ and $f_a^2 = \mathcal{O}(1/N_f)$, so $c_a$ evolve at a different time scale than $f_a^2$. This means that the evolution of the $c_a$
and the $f_a$ decouple, the $f_a$ being fast variables while the $c_a$ (as well as $\Lambda$ as we shall see) evolve adiabatically w.r.t. these fast variables.

Our next goal is to derive macroscopic equations associated to this dynamical system.  
Before that let us discuss qualitatively the behavior of the microscopic equations.
First notice that equations~(\ref{eq:cat2}) tend clearly to align the population matrix with the signal in the sense discussed in Section~\ref{sec:LambdaEff} for having
good generalization. Indeed we see that the rate of increase of $c_a$ is proportional to $f_a^2$, hence modes well aligned with the signal grow faster. 
The second equation is also driving the system into the good direction, regarding generalization.
The first term on the rhs of equation~(\ref{eq:fat2}) is
a positive source term corresponding to increasing the specification of the feature w.r.t. the signal,
i.e. to increase $r$ as we shall see in Section~\ref{sec:macro}. Because of the factor $1/(1+\Lambda c_a)$ we see that this increase goes preferentially to the weak modes. The second term instead   
conserves the longitudinal spectral power of the signal wrt the features but it induces a net flow of spectral power from weak to strong modes.
When combined these mechanisms contribute to improve the alignment of the population matrix with the signal.

\subsection{Effect of the architecture}\label{sec:constraints}
So far we did not take into account the limits and biases induced by the specific architecture of the model like for instance if the feature matrix
$F(\theta)$ is obtained from the last layer of a deep neural network. Considering this effect would require to project equations~(\ref{eq:dotomega})
on the deformation axes favored by the architecture. 
The generators of rotation have indeed the following decomposition
\[
d\Omega_{ab} = \sum_k \dot \Omega_{ab}^k (\theta) d\theta_k 
\]
where $\Omega^k$ is the rotation induced by an infinitesimal variation of the $k$th component of the parameter vector. This does span only a subset of all possible rotations,
and it is determined solely by the architecture of the model in a complicated way.
In the end this would result in some damping factor on the evolution of the $c_a$ in~(\ref{eq:cat2}),
and in some discount factors applied on the rotation generators especially on the one exchanging longitudinal ($a\le N_f$) with transverse vectors ($b>N_f$).
This latter effect is important as it corresponds to the inductive bias of the architecture which would for instance favor smooth functions and penalize rotation toward rugged ones. 
A minimal way to take this into account is to modify the ridge penalty term in~(\ref{eq:loss2}). By default is was choosen isotropic,
but we may replace it by some structured one 
\begin{equation}\label{eq:loss3}
\LL(\w,\theta) = \frac{1}{2}\int d\x d\x' \w^t F(\x\vert\theta)^tS(\x,\x')F(\x'\vert\theta)\w +\frac{\alpha}{N_s}\sum_{s=1}^{N_s}\bigl[y^{(s)}-\w^tF(\x^{(s)}\vert\theta)\bigr]^2,
\end{equation}
where $S$ is a positive definite kernel function. For instance if the input space
is a functional space spanned by a Fourier basis, $S$ could be diagonal in this basis with some dependence in $\vert\bk\vert^2$ with respect to wave numbers.
This would add extra terms in the microscopic dynamical equations that we won't investigate in the present work. Keeping the ridge penalty term in isotropic
form is already partially taking care of the implicit regularization of the model. Other arguments given in~\cite{louart2018random}
provide some justification for such considerations.
Additionally in order to justify that  $F$ can be considered as a matrix and derived components-wise
requires to consider the system to be in a  ``semi-lazy'' regime as explained on Figure~\ref{fig:student}, by reference to the lazy regime discussed in Section~\ref{sec:ntk}.
We assume only the parameter of the vector of feature function to be in the lazy regime, the weights of the last layer remaining arbitrary.
In such case we are send back to the problem of learning a $D\times N_f$ feature matrix, with $D$ being however potentially prohibitively large
as discussed in Section~\ref{sec:ntk}, depending on the chosen level of resolution when $d\gg 1$. There additional considerations have to come into play to make sense of an efficient
learning dynamics in such cases, because $r(t=0)$ would be otherwise very close to zero at initialization and the learning would have a hard time to start up.
These considerations like intrinsic dimension of the input $\x$ much lower than $d$, 
the inductive bias of the NN or weight sharing due to the composite nature of the input $\x$~\cite{lin2017does} have been discussed mostly qualitatively but could be
formalized in present framework by considering a feature matrix with some structure and symmetries. This is however outside the scope of the present paper.

\subsection{Macroscopic equations}\label{sec:macro}
The equations~(\ref{eq:cat2},\ref{eq:fat2}) determine completely the learning dynamics at the microscopic level in absence of architecture constraints.
At the macroscopic level, the dynamics would be conveniently summarized by those of $g(t)$ and $g_2(t)$ (or equivalently $\Lambda(t)$ and $\gd(t)$), $\mu_1(t)$, $\mu_2(t)$  and $r(t)$.
Unfortunately, as we shall see we cannot obtain in general a closed system of equations. Instead as usual in statistical
physics, the macroscopic description is obtained in the form of a hierarchy of equations involving all the moments of some distribution.
At least the dynamics of $r$ is straightforward, it is directly obtained from~(\ref{eq:fat2}) by summing the $f_a\dot f_a$. We get 
\begin{equation}\label{eq:rdot}
\dot r = 2\Lambda^2 r(1-r-\sigma^2)\mu_1,
\end{equation}
with again $\mu_1\in [0,1]$ given by~(\ref{def:mu_k}) ($\mu^\parallel = r$).
This yields a sigmoid-like behaviour:
\[
r(t) = \frac{1-\sigma^2}{1+C\exp\Bigl(-2\int_0^t d\tau \Lambda^2(\tau)\mu_1(\tau)\Bigr)},
\]
with $C>0$ a constant determined by initial conditions. In order to get the dynamics of $\Lambda$, we simply combine~(\ref{eq:dLambdaca}) with(\ref{eq:cat2})
to obtain
\begin{equation}\label{eq:Lambdadot}
\frac{d}{dt}\bigl[\Lambda(t)^{-1}\bigr] = \frac{2\Lambda r}{N_f}\frac{\tilde \mu_3-\tilde \mu_4}{\rho-\gd},
\end{equation}
with $\gd$ given by~(\ref{eq:QLambda}) and
\[
\tilde \mu_k \egaldef \mu_k+\frac{\mu_2}{\rho-\mathcal{Q}}g_k.
\]
Since $\mu_3>\mu_4$ and $g_3>g_4$ and $\rho > \mathcal{Q}$,
we see that the effective ridge penalty $\Lambda^{-1}$, or equivalently the test to train error ratio, is strictly increasing with time,
leading eventually, as expected, to overfitting. But due to the $1/N_f$ factor, we see that its dynamics is slow compared to those of $r$. 
In order to close the equations we need also the derivatives of $g_k$ and $\mu_k$.
For the $g_k$ we get
\begin{equation}\label{eq:gkdot}
\dot g_k = \frac{2kr\Lambda^2}{N_f}\Bigl[\tilde\mu_{k+3}-\tilde\mu_{k+2}+\frac{(\tilde\mu_3-\tilde\mu_4)(g_k-g_{k+1})}{\rho - \mathcal {Q}}\Bigr],
\end{equation}
the $1/N_f$ in front indicates that these are slow variables.
Concerning the $\mu_k$ instead, their dynamics which is dominated by the variations of $f_a$, decouple from the previous ones.
Neglecting the contributions from the variation of $c_a$ and $\Lambda$ of order $1/N_f$, and exploiting the anti-symmetry of the second r.h.s. term in~(\ref{eq:fat2})
we obtain 
\begin{equation}
  \dot \mu_k = 2\Lambda^2\Bigl[(1-r-\sigma^2)(\mu_{k+1}-\mu_1\mu_k) +  r\sum_{q=0}^{k-1}(\mu_{q+2}-\mu_{q+1})\mu_{k+1-q}\Bigr]\label{eq:mukdot}.
\end{equation}
This last equation is valid as long as the spectral density of the signal do not condensate on a mode or on a subset of degenerate modes, otherwise additional terms need
to be taken into account. 
The problem is that in this form this equation leads to a non-physical stationary state. The first term on the rhs is going to saturate at some
point when $r$ approaches its limit $1-\sigma^2$ while the second one is strictly negative and vanishes when all the $\mu_k\to 0$.
Coming back to the discussion of equation~(\ref{eq:fat2}), it is actually already expected that the microscopic dynamics leads to concentrate the spectral power of the signal  on the
strong modes, eventually leading to the apparition of a condensate on the strongest one. To take this into account we need to modify slightly the equations
which will be done in the next section.
The system (\ref{eq:rdot},\ref{eq:mukdot},\ref{eq:Lambdadot},\ref{eq:gkdot}) constitute a hierarchical set of macroscopic equations describing the learning process in closed form.
Actually it is decomposed into two subsystems   (\ref{eq:rdot},\ref{eq:mukdot}) and (\ref{eq:Lambdadot},\ref{eq:gkdot}) respectively of fast and slow variables,
which means that in practice to solve it one should insert at each time step of the slow system,
the stationary state of (\ref{eq:rdot},\ref{eq:mukdot}) conditionally on $\Lambda$ and the $g_k$'s.

\subsection{Macroscopic dynamics with condensate}\label{sec:condensate}
The equations derived so-far from RMT assume in particular that the signal decomposes along the modes of the population matrix with components of order $f_a = \mathcal{O}(1/\sqrt{N_f})$.
Consequently we assume that the signal condensate on a degenerate level $c_a$ with degeneracy $\mathcal{O}(N_f)$ to stay within valid hypothesis regarding RMT.
As a result equations~(\ref{eq:cat2},\ref{eq:Lambdadot}) correspond still to a slow dynamics. 
Let $c_{\rm max}$ the level associated to the condensate and $\mu^c$ the   longitudinal spectral power fraction of the signal on the condensate.
Hence  we have the decomposition
\[
\mu^\parallel(x)  = \mu^{\rm bulk}(x) + \mu^\parallel\mu_{\rm cmax} \delta(x-c_{\rm max}).
\]
Then we can write accordingly the following decomposition
\[
  \mu_k = \mu_k^{\rm bulk}+\mu_{\rm cmax} y^k,
\]
where
\[
y \egaldef \frac{1}{1+\Lambda c_{\rm max}}.
\]
Including the term which was neglected in~(\ref{eq:mukdot}) yields
\[
  \dot \mu_k = 2\Lambda^2\Bigl[(1-r-\sigma^2)(\mu_{k+1}-\mu_1\mu_k) + r\sum_{q=0}^{k-1}(\mu_{q+2}-\mu_{q+1})\mu_{k+1-q}
  +k r\bigl(\mu_{k+2}^{(2)}-\mu_{k+3}^{(2)}\bigr)\Bigr],
\]
where
\[
\mu_k^{(2)} \egaldef \frac{1}{r^2}\sum_{a=1}^{N_f}\frac{f_a^4}{(1+\Lambda c_a)^k}.
\]
Here the index $a$ may represent indifferently individual modes or more conveniently spectral level associated with $c_a$. In the latter case  
$f_a^2$ represents the spectral density on this level which can be of order $\mathcal{O}(1)$ instead of $\mathcal{O}(1/N_f)$ if the degeneracy of that level is extensive.
This is the case for the condensate which therefore gives the contribution
\[
\mu_k^{(2)} \egaldef \mu_{\rm cmax}^2y^k+\mathcal{O}(1/N_f).
\]
Finally we get 
\begin{equation}
  \dot \mu_k = 2\Lambda^2\Bigl[(1-r-\sigma^2)(\mu_{k+1}-\mu_1\mu_k) +  r\sum_{q=0}^{k-1}(\mu_{q+2}-\mu_{q+1})\mu_{k+1-q}
  +rk\mu_{\rm cmax}^2 (1-y)y^{k+2}\Bigr].\label{eq:mukcondot}
\end{equation}
Closing this system requires the introduction of additional macroscopic variables in order to express the dynamics of $\mu_{\rm cmax}$ leading to cumbersome equations.
Instead we will content ourselves with a simple bounding of the derivative of $\mu_{\rm cmax}$. From the microscopic equation~(\ref{eq:fat2}) we have
\[
\dot\mu_{\rm cmax} = \Lambda^2y\mu_{\rm cmax}\Bigl[1-r-\sigma^2+\sum_{b=1,c_b\ne c_{\rm max}}^{N_f}\frac{c_{\rm max}+c_b}{c_{\rm max}-c_b}\frac{f_b^2}{1+\Lambda c_b}\Bigr],
\]
Both terms on the rhs are positive. Let $\Delta = c_{\rm max}-\max_{b,c_b\ne c_{\rm max}}(c_b)$
the spectral gap between the condensate and the bulk. We have
\[
\Lambda^2 y\bigl[\mu_1^{\rm bulk} +(1-r-\sigma^2)\bigr] \le \frac{\dot\mu_{\rm cmax}}{2\mu_{\rm cmax}} \le 
\Lambda^2 y\bigl[\frac{2 c_{\rm max}-\Delta}{\Delta}\mu_1^{\rm bulk} +(1-r-\sigma^2)\bigr].
\]
From  this it is now clear that $\mu_{\rm cmax}$ increases as long as $r<1-\sigma^2$ and $\mu_1^{\rm bulk} > 0$, i.e. $\mu_{\rm cmax}<1$.
As a result, as can be checked, equation~(\ref{eq:mukcondot}) has a stationary state corresponding to $\mu_{\rm cmax} = 1$ i.e. $\mu^{\rm bulk}=0$, and $r=1-\sigma^2$, meaning that the signal
is completely condensed on the strong modes, which as discussed in Section~\ref{sec:LambdaEff} is a crucial point for obtaining good generalization
properties. In practice it is not clear whether the dynamics can maintain a degenerated set of strong modes, or if instead isolated modes appear at the top of the
spectrum. In that case the dynamics of previously slow variables become faster
(because $f_a^2$ in equation~(\ref{eq:cat2}) is $\mathcal{O}(1)$) and coupled to the dynamics of the $\mu_k$, which should result in a more complex behavior.
Since also the validity of RMT in this case is not  granted we will leave this aside and focus instead on simple
solvable special cases in Section~\ref{sec:Specialcases2}.
\begin{figure}
  \centerline{\includegraphics[width=0.5\textwidth]{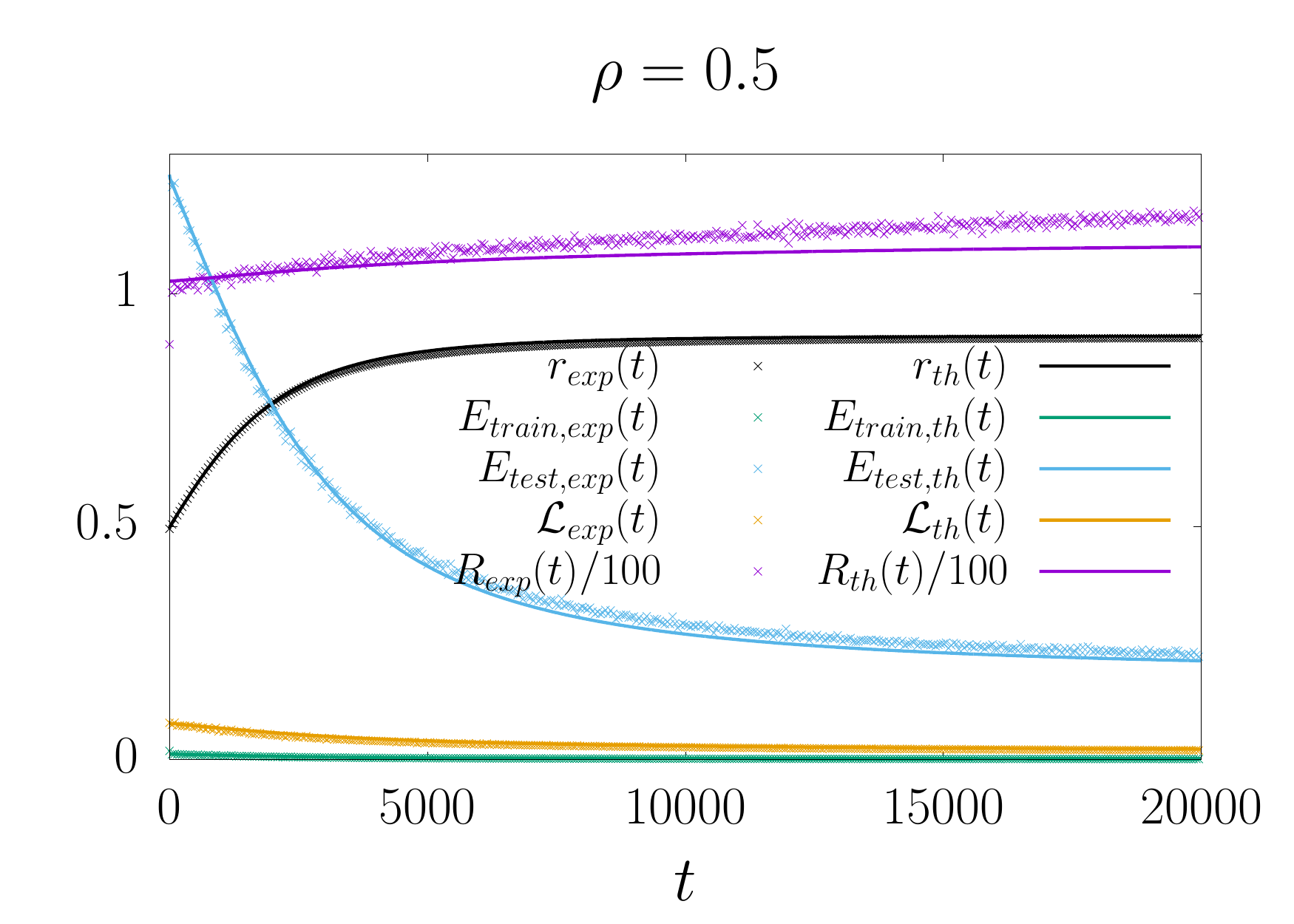}
  \includegraphics[width=0.5\textwidth]{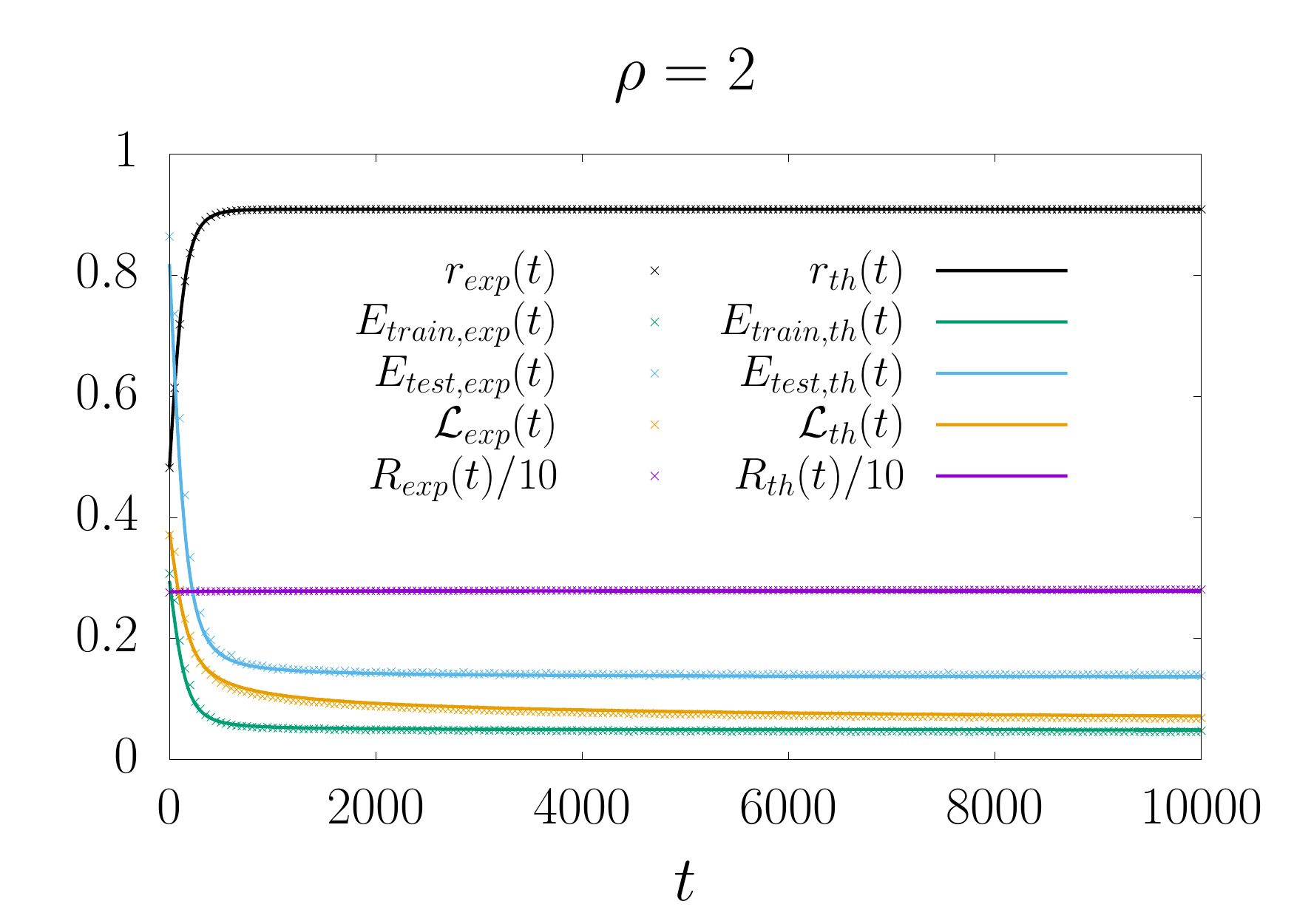}}
  \centerline{\includegraphics[width=0.5\textwidth]{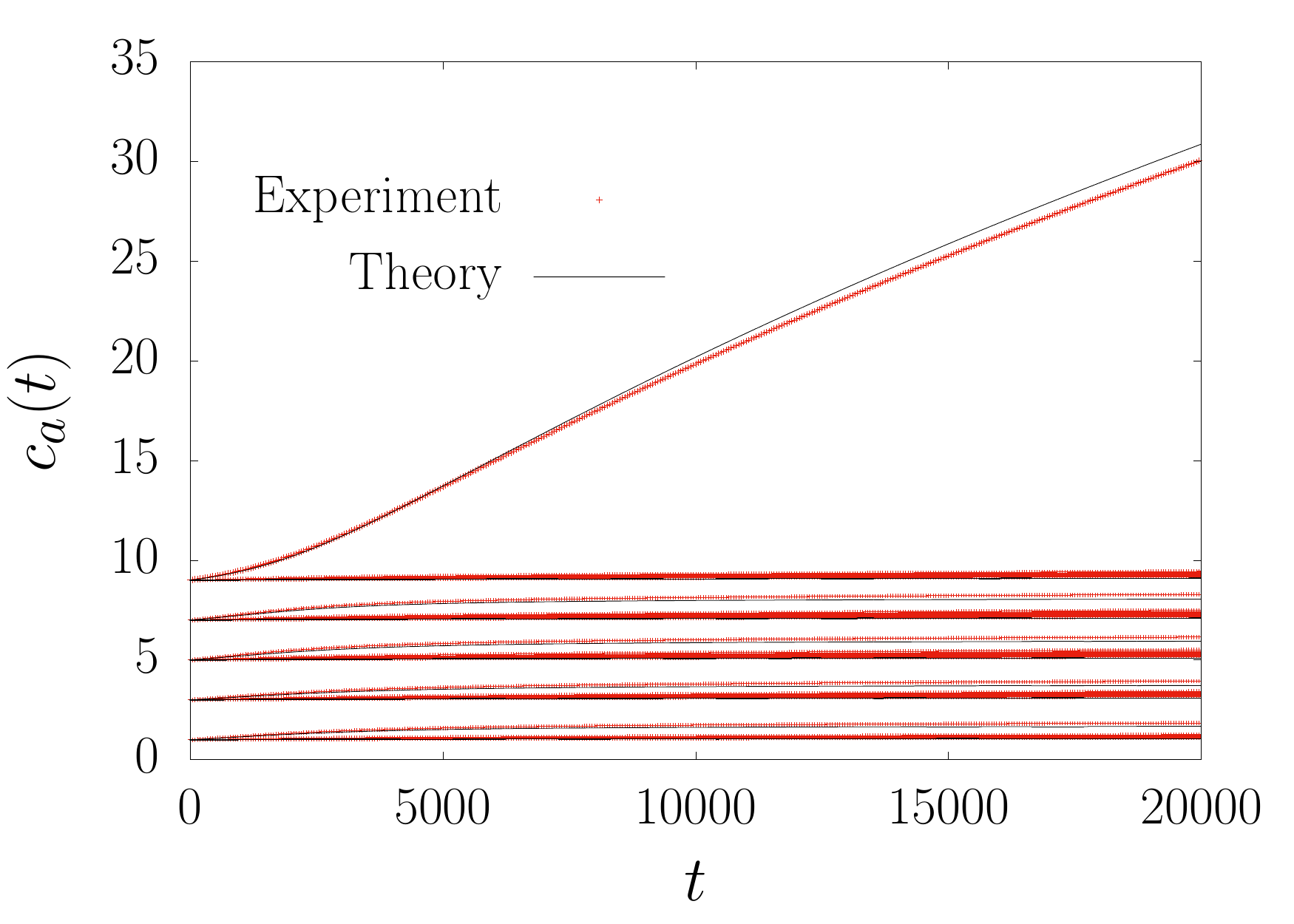}
  \includegraphics[width=0.5\textwidth]{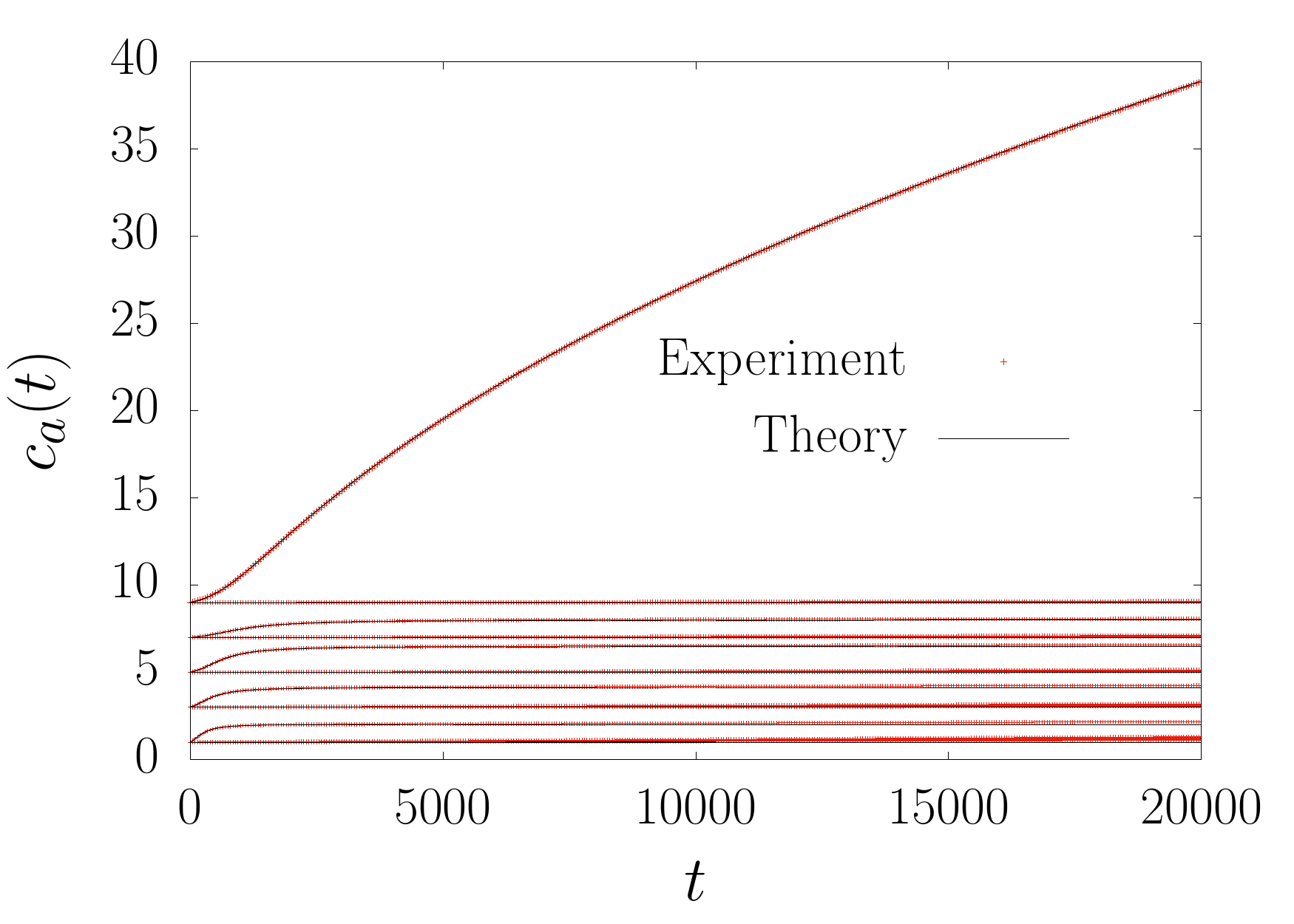}}
  \centerline{\includegraphics[width=0.5\textwidth]{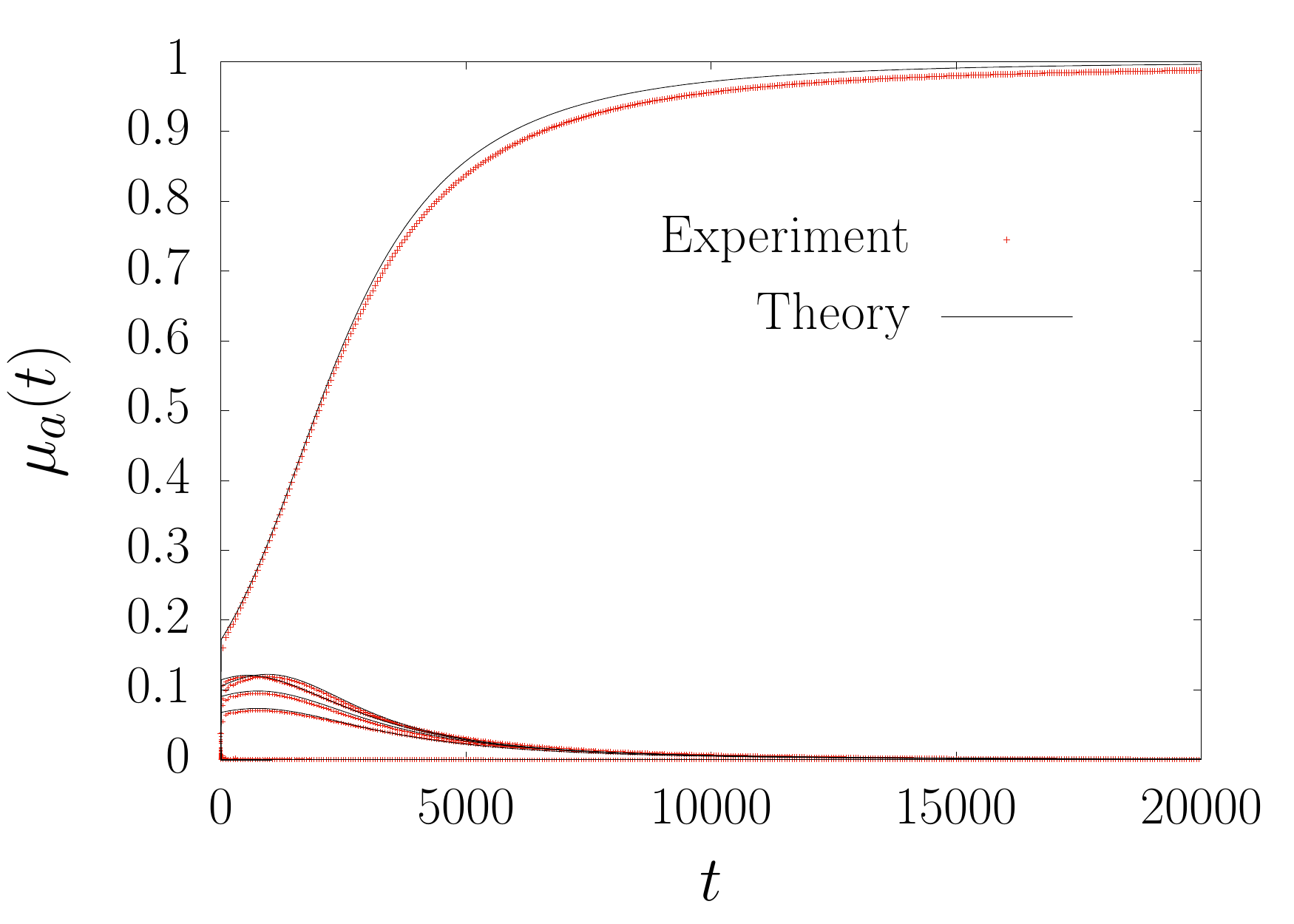}
  \includegraphics[width=0.5\textwidth]{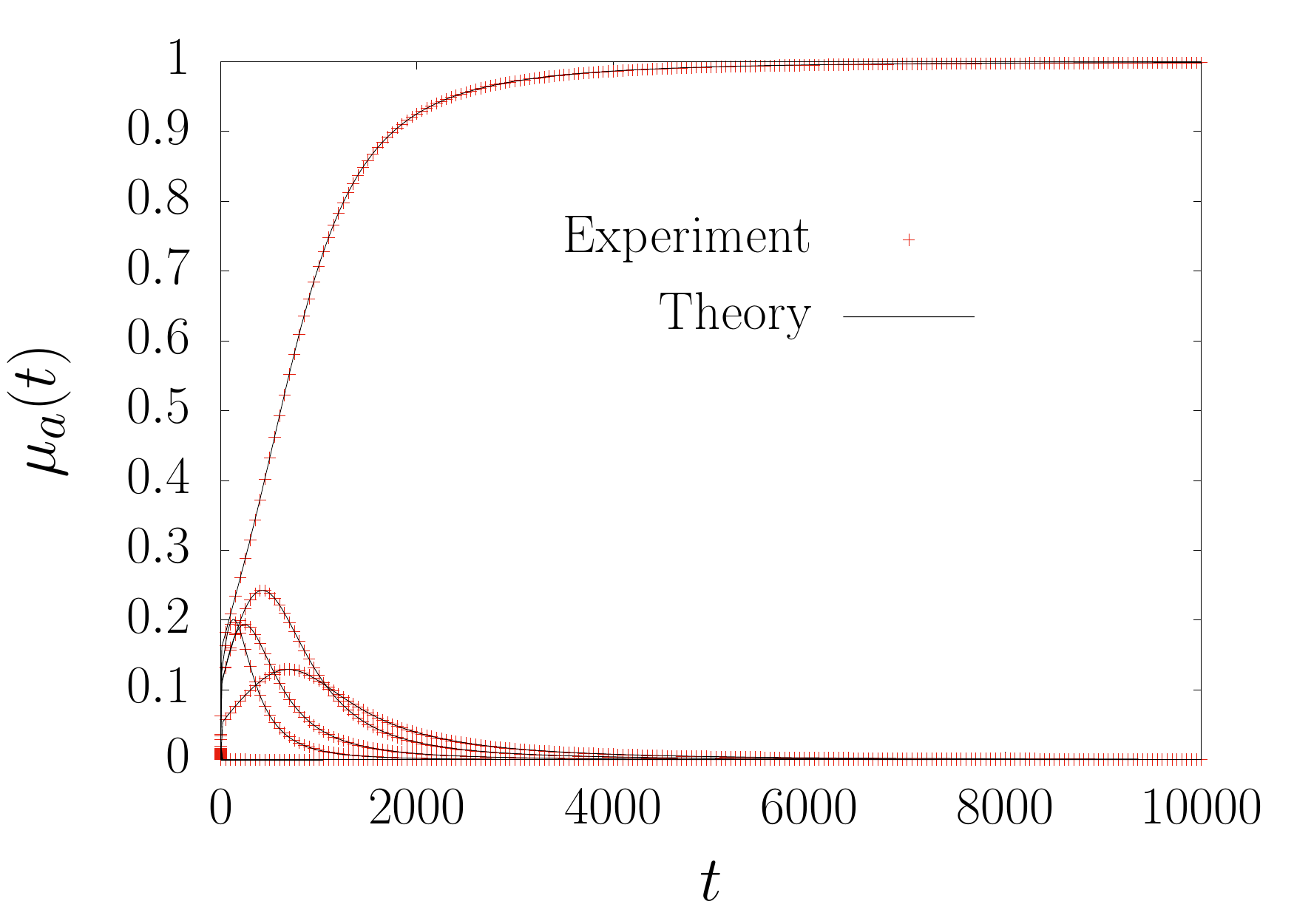}}
  \caption{\label{fig:dynamics}
    Feature learning process dynamics: comparison between actual feature learning processes and the corresponding dynamical system (\ref{eq:cat2},\ref{eq:fat2}). The left and right
    panels are respectively in the over and under-parameterized regime with same level of noise $\sigma^2=0.1$.
    The population matrix is initialized with $5$ equally spaced and degenerate levels. Concerning the learning process we have in both cases $N_f = D/2 = 100$.}
\end{figure}

On Figure~\ref{fig:dynamics} is shown a comparison between the actual gradient descent and the autonomous dynamical system~(\ref{eq:cat2},\ref{eq:fat2}),
which actually confirms that condensation occurs.
The dynamical equations are written with the hypothesis that the train data are un-correlated from one time step to the next one, which is not necessarily true
in practice because in general the training set is finite, except for streaming data. To keep with a consistent definition of $\rho$
that represents the ratio of training data to the number of parameters of the model the experimental setting is the following: to each time step is
associated a batch of training data of size $N=\rho N_f$ out of a training set of overall size 
$N_{\rm train}\egaldef\rho\times N_f(D+1)$, where $N_f(D+1)$ represent the total number parameters of the model, $N_f$ being the number of weights of the last layer and
$D N_f$ being the number of entries in the feature matrix $F$.
The dynamical system is initialized with
the eigenvalues of the population matrix and the components of the signal on the corresponding 
eigenmodes given at the initialization of the experiment. On the examples of Figure~\ref{fig:dynamics} 
the population matrix is initialized with $5$ equally spaced and degenerate levels and to compare we monitor the eigenvalues $c_a(t)$, the components $f_a(t)$
and various macroscopic variables of interest like train and test errors $E_{\rm train,test}(t)$, the loss $\LL(t)$,
the longitudinal power $r(t)$ of the signal and the train-test error ratio ${\mathcal R}^2(t)$. For the actual learning experiment, at each time step,
$E_{\rm train}$ as well as $\LL$ are evaluated on the batch of training data, while $E_{\rm test}$ is evaluated with $10^4$ independent test samples and
${\mathcal R}^2(t)$ is evaluated via the empirical estimation $\hat\Lambda$ of $\Lambda$ given in Section~\ref{sec:practical}. For the dynamical system,
equations~(\ref{eq:Etrainasymp2},\ref{eq:Etestasymp},\ref{eq:Lossasymp},\ref{def:Lambdaeff})
are used along with the fixed point equation~(\ref{eq:FP}) for $\Lambda$ and the finite size counterparts~(\ref{eq:gk-muk}) of the spectral coefficients~(\ref{def:g_k},\ref{def:mu_k}). 
As we see, for $\rho=2$, i.e. in the under-parameterized regime the autonomous system reproduces exactly the dynamics of the experiments, remarkably 
in a situation outside of the domain of validity of the equations, because some condensate emerge from each levels, temporarily for all except the top level.
In the over-parameterized regime, i.e. $\rho=0.5$ the dynamics is also very-well reproduced, but we see a small deviation as time progresses due to the finite
size of the training data set. This is to be expected from the finite size of the training set, 
the deviation becoming more pronounced for small values of $\rho$, but indeed disappear (not shown) when the experiment is feed with un-correlated streaming-data. 
In order to understand this consider the ratio
\[
\rho_D \egaldef \rho\frac{N_f(D+1)}{D},
\]
which represents the number of training data points per direction in the embedding space. Its value is respectively $50$ and $200$ for $\rho=0.5$ and $\rho=2$ in the experiments.
Each input vector $\x_s$ being normalized to $\sqrt{D}$, the empirical operator
\[
\hat\I = \frac{1}{N_{\rm train}}\sum_{s=1}^{N_{\rm train}}\x_s\x_s^t \Lra_{N_{\rm train}\to\infty} \I
\]
is a finite $N_{\rm train}$ approximation to the identity operator in the embedding space and the vector
\[
\tilde f \egaldef \hat\I f
\]
is the signal as viewed from the training set. From RMT we have
\[
\Vert\f-\tilde\f\Vert^2 \approx \frac{\Vert \f\Vert^2}{\rho_D},
\]
and the training process is going to point to $\tilde f$ instead of $\f$ which qualitatively explain the observed deviation.
Since $\rho_D\propto N_f$ this is however a finite size effect which vanishes in the thermodynamics limits.

\subsection{Special cases}\label{sec:Specialcases2}
Let us consider some cases where the macroscopic equations can be closed. These correspond to a population matrix having a finite number of degenerate levels.
According to the microscopic equations, all the modes of the same level evolve the same way in that case, provided that the spectral power of the signal
is equally distributed at initialization among the modes belonging to the same levels (identical $f_a^2$). Here of course it is assumed that no spontaneous symmetry breaking is taking place.
What is shown on Figure~\ref{fig:dynamics} proves actually the contrary,
this instability is observed and well explained by equation~(\ref{eq:fat2}) which tends to amplify differences between longitudinal modes.
Anyway it is always possible to consider a scenario where degenerate levels of the population matrix are maintained artificially
during the learning, by mixing the modes and averaging the eigenvalue within each level after each iteration. This gives consistent results (not shown) 
where the special cases considered here become relevant.

\paragraph{The one-level case:} is the first obvious case and corresponds to have a population matrix with one single degenerate level.
So again in this case we have $g_k = \mu_k = y^k$ with $y\in [\max(0,1-\rho),1]$,
so that the dynamical equation~(\ref{eq:Lambdadot}) concerning $\Lambda$
can be rewritten directly in term of $y$ and $r$ only and we end up with two coupled equations:
\begin{align}
  \dot y &= -\frac{2\alpha^2 r}{\rho^2 N_f}\frac{(y+\rho-1)^3}{\rho-(1-y)^2}y^3(1-y)\Bigl[1+\frac{y^2}{\rho-(1-y)^2}\Bigr],\label{eq:dg}\\[0.2cm]
  \dot r &= \frac{2\alpha^2}{\rho^2} r(1-r-\sigma^2)y(\rho-1+y)^2.\label{eq:dr}
\end{align}
First $r$ which is a fast variable tends to $1-\sigma^2$, then $y$ at a slower speed will eventually tends to $0$.
Note that equation~(\ref{eq:mukcondot}) is irrelevant in this case, as can be checked we get $\dot \mu_k=0$ which corresponds to absence of fast variations
of $\mu_k$. At large time we therefore get the generalization error given by~(\ref{eq:Etestsingle}) with $g=g_{\rm min}$
which actually corresponds to the overfitting regime ($\Lambda=0$). In order to obtain the best result, one should stop in that case the optimization when
$g=g^\star$ given in~(\ref{eq:gstar}) with $r=1-\sigma^2$ which maybe unknown actually.

\paragraph{The two-level case:} is more interesting and will be useful to illustrate the reinforcement mechanism on the strong modes.
As before we denote by $y_a \egaldef 1/(1+\Lambda c_a)$ for each level $a\in{1,2}$. We have the decomposition
\begin{align*}
g_k &= (1-\nu) y_1^k+\nu y_2^k, \\[0.2cm]
\mu_k &= (1-\mu) y_1^k+\mu y_2^k,
\end{align*}
where $\nu$ is the fraction of strong longitudinal modes which remains fixed, while $\mu\in[0,1]$ is the fraction of the spectral power of the signal
aligned with the features which are contained in the strong level;
Consequently the effective coupling reads
\begin{equation}\label{eq:lambday1y2}
\Lambda(y_1,y_2) = \frac{\alpha}{\rho}\bigl[\rho-1+(1-\nu) y_1+\nu y_2\bigr].
\end{equation}
For $\rho \ge 1$ the valid domain ${\mathbb D}$ of these variable is $(y_1,y_2)\in [0,1]^2$. Instead for $\rho <1$, the constraint $g\in[1-\rho,1]$ (see Section~\ref{sec:LambdaEff})
result in ${\mathbb D}$ being the intersection of $[0,1]^2$ with
\begin{equation}\label{eq:constraint}
(1-\nu) y_1+\nu y_2 \ge 1-\rho,
\end{equation}
i.e. above the line
\[
\nu y_2 = (\nu-1)y_1+1-\rho,
\]
where $\Lambda=0$ in the $(y_1,y_2)$ plane.
As already explained in the last section $\mu$ is a fast variable, hence if $N_f$ is large enough
its dynamics is decoupled from the slow one and we can directly assume $\mu= 1$. It remains to examine the slow dynamics, which is given by
the evolution of $\Lambda$. We have
\[
\dot \Lambda = -y_2^2\frac{2 r}{N_f}\frac{\Lambda^3}{\rho-\gd}\Bigl[y_2^3(1-y_2)+\frac{g_3-g_4}{\rho-\gd}\Bigr]
\]
where the specification on the present case yields
\begin{align}
\gd &= (1-\nu)(1-y_1)^2 + \nu (1-y_2)^2,\label{eq:gd2}\\[0.2cm]
g_3-g_4 &= (1-\nu) y_1^3(1-y_1) + \nu y_2^3(1-y_2)^2.\label{eq:dg34}
\end{align}
In addition we have
\[
\dot y_a = -y_a(1-y_a)\bigl[\frac{\dot \Lambda}{\Lambda} + \frac{\dot c_a}{c_a}\bigr]
\]
and from~(\ref{eq:cat2})  we have
\begin{align*}
\frac{\dot c_1}{c_1} &= \frac{2\Lambda^2 r}{N_f} \frac{y_1^2y_2^2}{\rho-\gd},\\[0.2cm]
\frac{\dot c_2}{c_2} &= \frac{2\Lambda^2 r}{N_f} y_2^2\Bigl(\frac{1}{\nu}+ \frac{y_2^2}{\rho-\gd}\Bigr).
\end{align*}
with $f_{a}^2 = r\bigl[N_f\nu\min(1,\rho)\bigr]^{-1}$ for modes in the strong level ($f_a^2=0$ for the other ones).
\begin{figure}[ht]
\centerline{\includegraphics[width=0.6\textwidth]{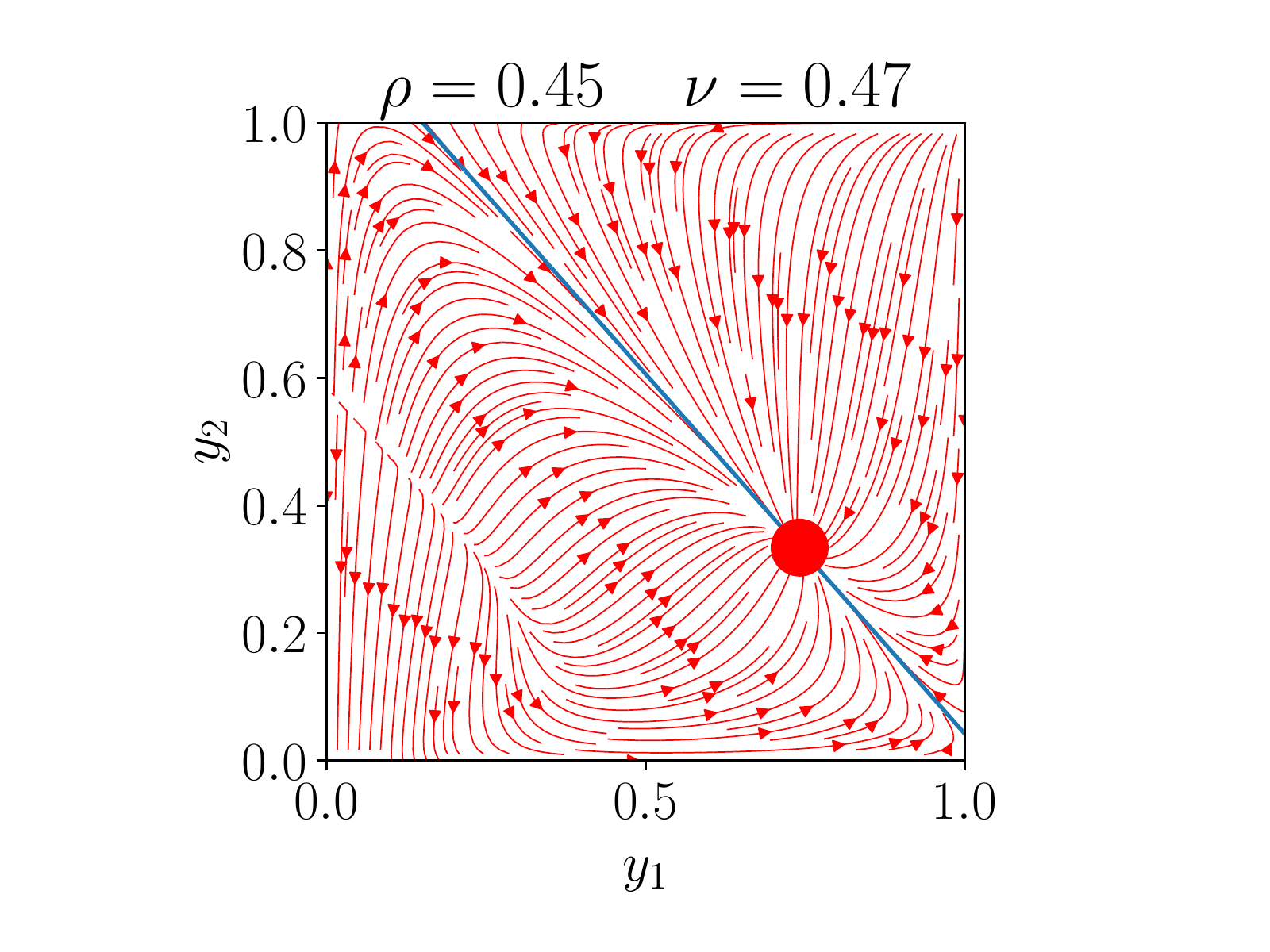}
  \includegraphics[width=0.6\textwidth]{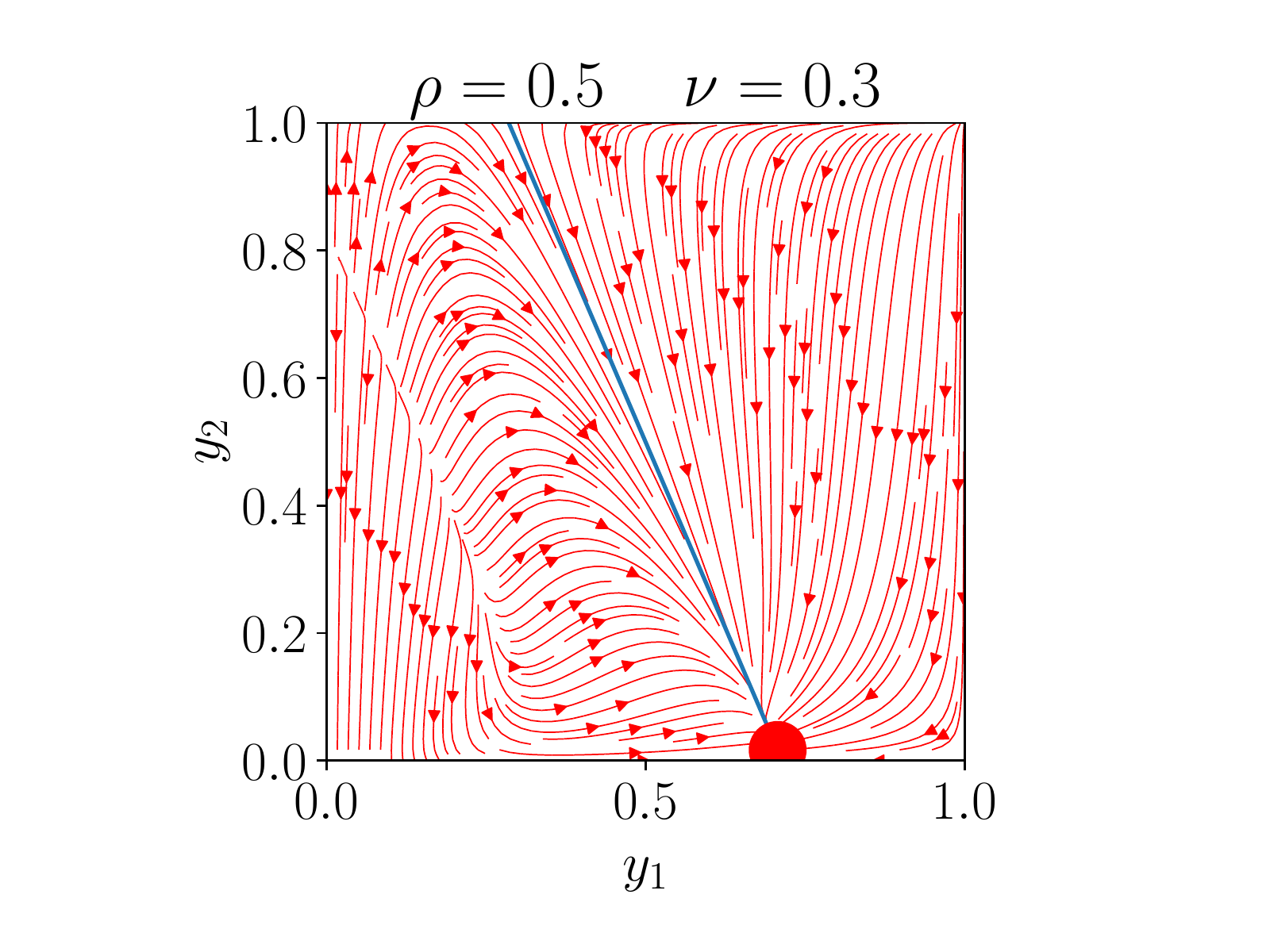}}
\centerline{\includegraphics[width=0.6\textwidth]{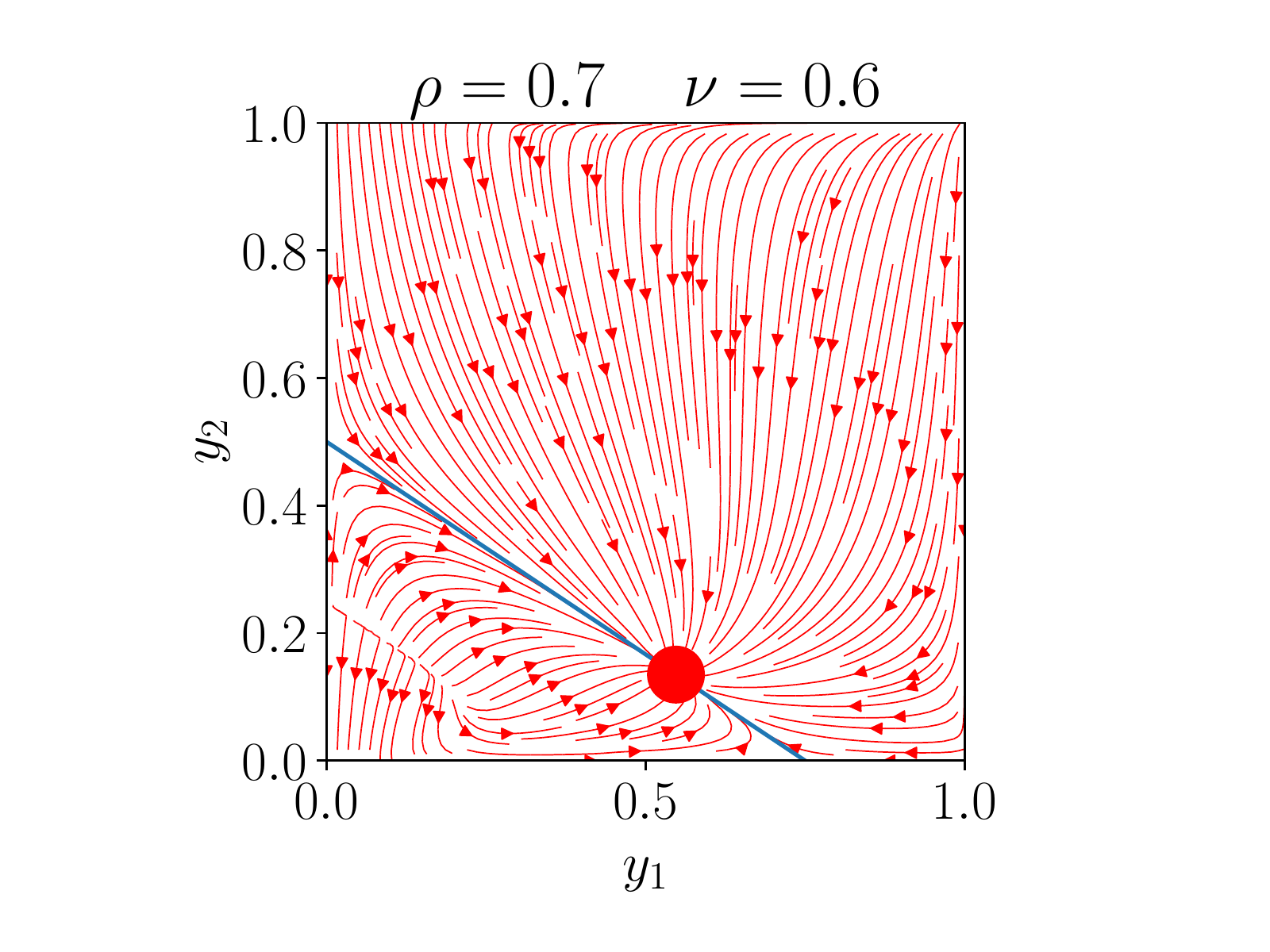}
\includegraphics[width=0.6\textwidth]{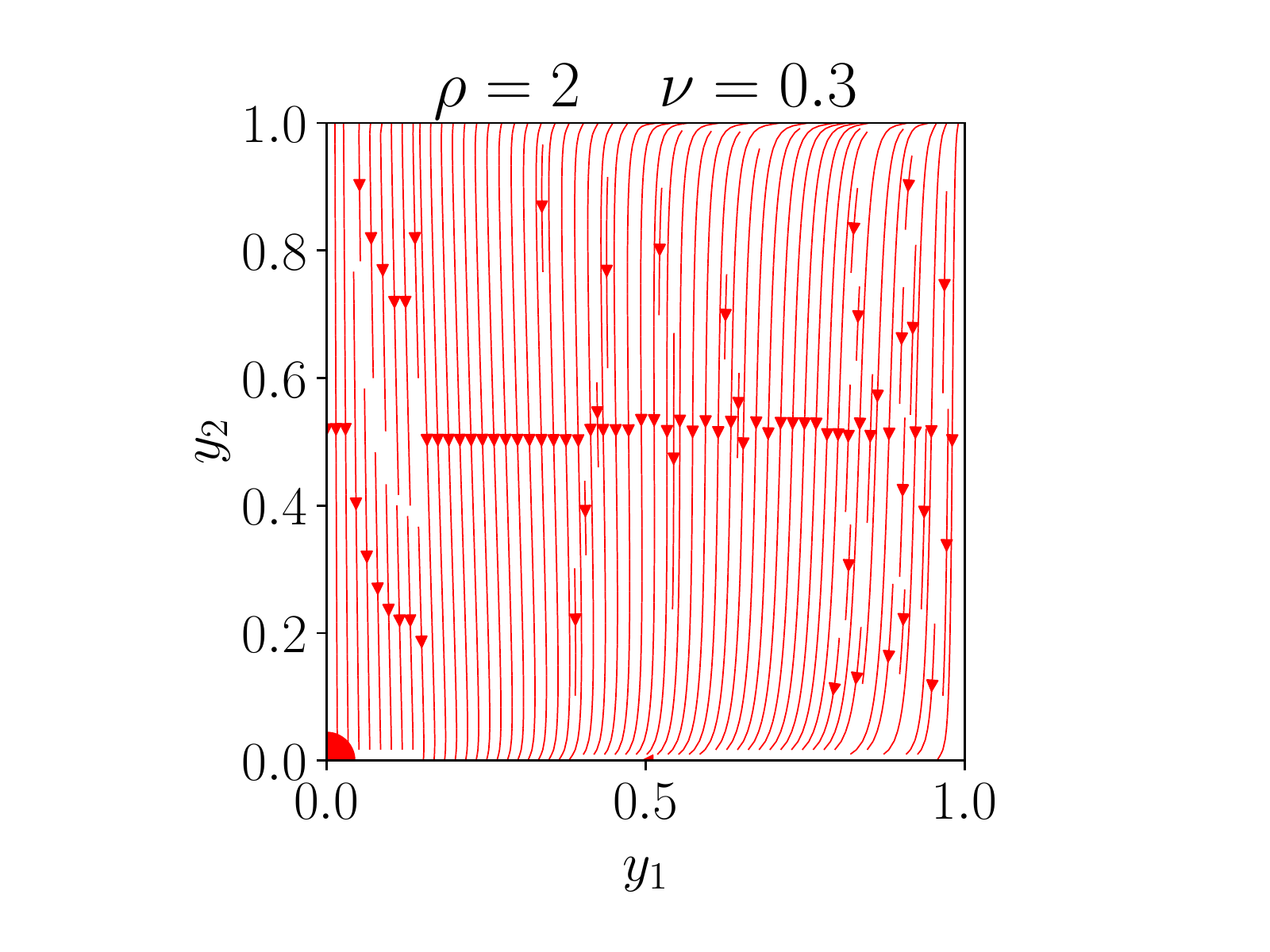}}
\caption{\label{fig:phase-portraits} Examples of phase portraits corresponding to the autonomous system~(\ref{eq:dy1},\ref{eq:dy2}). The red dot locate the stable
fixed point while the blue line represents $\Lambda(y_1,y_2)=0$ given by~(\ref{eq:lambday1y2}) above which is the valid domain ${\mathbb D}$.}
\end{figure}
Finally we obtain
\begin{align}
  \dot y_1 &= -\frac{2\Lambda^2 r}{N_f}\frac{y_1(1-y_1)y_2^2}{\rho -\gd}\Bigl(y_1^2-y_2(1-y_2)-\frac{g_3-g_4}{\rho-\gd}\Bigr),\label{eq:dy1}\\[0.2cm]
  \dot y_2 &= -\frac{2\Lambda^2 r}{N_f}y_2^3(1-y_2)\Bigl(\frac{1}{\nu}+\frac{y_2(2y_2-1)}{\rho-\gd}-\frac{g_3-g_4}{(\rho-\gd)^2}\Bigr),\label{eq:dy2}
\end{align}
which altogether with~(\ref{eq:gd2},\ref{eq:dg34}) and (\ref{eq:lambday1y2})
defines an autonomous system for the variables $(y_1,y_2)$. 
In terms of $(y_1,y_2)$ the test error reads
\[
E_{\rm test}(y_1,y_2) = \frac{\rho\bigl[1+r(y_2^2-1)\bigr]}{\rho-(1-\nu)(1-y_1)^2-\nu(1-y_2)^2}.
\]
Some phase portraits of this simple dynamical system are shown on Figure~\ref{fig:phase-portraits} with different combinations of parameters $\rho$ and $\nu$. 
As can be easily checked, the point $(y_1,y_2) = (0,0)$ is a stable fixed point of this system which is in the valid domain ${\mathbb D}$ defined above only for $\rho>1$.
Other stationary points like $(y_1,y_2) \in \{(1,0),(0,1),(1,1)\}$ are unstable.
For $\rho>1$ no other fixed points  exit and the system always converges to $(0,0)$.
For $\rho\le 1$ this point is outside ${\mathbb D}$, but as can be shown, there are two additional possible fixed points: 
$(y_1,y_2) = \Bigl(\frac{1-\rho}{1-\nu},0\Bigr)$ and $(y_1,y_2)= \Bigl(\sqrt{1-\rho},\frac{1-\nu}{\nu}\bigl[\frac{1-\rho}{1-\nu}-\sqrt{1-\rho}\bigr]\Bigr)$, both lying
on the line $\Lambda(y_1,y_2) = 0$ but not necessarily in ${\mathbb D}$.
When $\rho\in[\nu(2-\nu),1]$ the first one is a stable fixed point, the second being outside ${\mathbb D}$, while for
$\rho\in[0,\nu(2-\nu)]$ the second is always the only stable fixed point. These four situations are illustrated on 
Figure~\ref{fig:phase-portraits}.  
The test error which is reached at these various fixed points is then given by
\[
E_{\rm test}(t\to\infty) =
\begin{cases}
  \DD \frac{\rho}{\rho-1}(1-r)\hspace{5cm} \rho\ge 1, \\[0.3cm]
  \DD \frac{\rho}{1-\rho}\frac{1-\nu}{\rho-\nu}(1-r)\hspace{4cm} \nu(2-\nu) \le \rho \le 1,\\[0.2cm]
  \DD E_{\rm test}\Bigl(\sqrt{1-\rho},\frac{1-\nu}{\nu}\bigl[\frac{1-\rho}{1-\nu}-\sqrt{1-\rho}\bigr]\Bigr)\qquad \rho <\nu(2-\nu).
\end{cases}
\]
None of these limit points are optimal, in particular close to $\rho=1$ the system is badly overfitting.
A best scenario is obtained with $\nu$ small and $(y_1,y_2)$ converging close to $(1,0)$ zero and $y_1$ close to one, which is
obtained with $\rho\simeq\nu(2-\nu)$. Overall we see that the long term dynamics is not necessarily beneficial and may lead to overfitting,
in contrary to the short term dynamics of the population matrix discussed in Section~\ref{sec:micro} and~\ref{sec:condensate}.

\section{Discussion}
The dynamical system~(\ref{eq:cat2},\ref{eq:fat2}) and associated macroscopic equations, obtained in the free probability regime
describe very precisely the feature learning process of an ideal NN, sufficiently large and expressive to be considered in a free of constraint and ``semi-lazy'' regime defined in
Section~\ref{sec:constraints}. Various dynamical features are revealed by this dynamical system. First the orientation of the population matrix with the signal
appears to occur at a faster scale than the dynamics of it eigenvalues. Both scale are separated by a factor $N_f$, i.e. the number of features to be learned.
In the first stage of the learning, 
information from the signal is mainly collected by the weak modes and then transferred to the strongest one which emerge spontaneously. This goes in the direction 
of having good generalization propertied of the solution, according the alignment criteria of the population matrix with the signal discussed around the general 
formula~(\ref{eq:Etestasymp}) of the test error derived in the free probability regime thanks to the train-test error ratio discussed in
Sections~\ref{sec:Test-train},\ref{sec:asymp} and~\ref{sec:LambdaEff}. Instead the long term dynamics offers less guaranties as seen with the dynamical
system~(\ref{eq:dy1},\ref{eq:dy2}) corresponding to a $2$-level artificially maintained population matrix.

These observations call for further investigations in different avenues, both practical and theoretical. From the practical point of view,
we may expect to be able to improve on feature learning algorithms, in order to insure that the result obtained is optimal for the generalization criteria
and that the time to reach the result is minimal. From the theoretical point of view, as suggested in Section~\ref{sec:constraints}, the
unreasonable success of deep learning~\cite{sejnowski2020unreasonable} could be investigated more quantitatively by rewriting the
dynamical equations with a structured feature matrix and appropriate hypothesis on the signal, and adapting them to the classification context instead of the
regression considered here. The possibility to transfer some observations of the present work to the unsupervised learning context seems also plausible thanks to
the observation made in a previous work that a convex relaxation of a restricted Boltzmann machine, called ``Coulomb machine'' could be learned in principle by means
of constraint linear regressions~\cite{DeFu2021}.

\subsection*{Acknowledgements}
I wish to thank Guillaume Charpiat for insightful discussions especially on neural networks.

\bibliographystyle{unsrt}
\bibliography{paper}

\appendix
\section{Random matrix theory and planar diagrams}\label{app:diag_expansion}
Let us first summarize without much details the original setting and derivation of the Marchenko-Pastur distribution~\cite{MaPa} adapted to our context.
We assume that $C$ is of the form
\[
C\n = C_0+\sum_{n=1}^N\omega_n\omega_n^t,
\]
where $C\n$ and $C_0$ are $N_f\times N_f$ symmetric matrices, and $\omega_n$ are random vectors of the form
\[
\omega_n = \Vert \omega_n\Vert \x_n 
\]
where $\x_n$ is a unit vector uniformly distributed on the $N_f$-dimensional sphere and the square norm $\Vert \omega_n\Vert^2$ of $\omega_n$ have the
same independent distribution $\sigma(x)$. We assume this to have a second moment $\tau$.
As a result the population matrix is given by
\[
C = C_0+{\mathbb E}\Bigl[\sum_{n=1}^N\omega_n\omega_n^t\Bigr] = C_0 + \rho\tau {\mathbb I},
\]
with $\rho = N/N_f$.
Let us call $\nu_\rho\n$ the spectral measure of $C\n$ (normalized to one) and $g\n(\rho,z)$ its Stieltjes transform:
\[
  g^{(\sn)}(\rho,z) = \int_0^\infty\frac{\nu_\rho^{(\sn)}(x)}{z+x} = {\mathbb E}\Bigl[\frac{1}{N_f}\Tr \Bigl(G\n(z)\Bigr)\Bigr],
\]
where
\[
G\n(z) \egaldef \bigl(z+C\n\bigr)^{-1}.
\]
In order to obtain the limit spectral density of $C\n$, we have to find the limit $g(\rho,z)$ of $g\n(\rho,z)$ when $N$ and $N_f$ tend to infinity at fixed $\rho$.
To this end  Marchenko and Pastur proceed by recurrence. Thanks to the Sherman-Morrison formula they indeed have 
\[
g\nn(\rho+d\rho,z) = g\n(\rho,z)-\frac{d\rho}{z+\omega_{\sn+1}^t G\n(z)\omega_{\sn+1}}\Tr\bigl(G\n(z)\omega_{\sn+1}\omega_{\sn+1}^t G\n(z)\bigr).
\]
with $d\rho = 1/N_f$. The limit $N\to\infty$ is taken at fixed $\rho$. Keeping $\Vert \omega_n\Vert^2=\tau_n$ fixed, under simple hypothesis
they show self-averaging properties of the resolvent
\begin{equation}\label{eq:c1}
\lim_{N\to\infty} \left\vert\frac{1}{N_f}\omega_{\sn+1}^t G\n(z)\omega_{\sn+1} - \tau_{\sn+1} g(\rho,z)\right\vert = 0
\end{equation}
and product of resolvents
\begin{equation}\label{eq:c2}
\lim_{N\to\infty} \left\vert\frac{1}{N_f}\Tr\bigl[G\n(z)\omega_{\sn+1}\omega_{\sn+1}^tG\n(z)\bigr] - \tau_{\sn+1}\Tr\Bigl[\bigl(z+C\n\bigr)^{-2}\Bigr]\right\vert = 0. 
\end{equation}
This then leads to get the following equation
\[
\frac{\partial}{\partial\rho} g(\rho,z) + \frac{\tau(\rho)}{1+\tau(\rho) g(\rho,z)} \frac{\partial}{\partial z} g(\rho,z) = 0,
\]
thanks to the identity
\[
\frac{1}{N_f}\Tr\Bigl(\bigl[z+C\n\bigr]^{-2}\Bigr) = \frac{\partial}{\partial z} g^{(\sn)}(\rho,z).
\]
This is solved via to the method of characteristics as
\[
g(\rho,z) = g\Bigl(0,z+\int_0^\rho dt \frac{\tau(t)}{1+\tau(t)g(\rho,z)}\Bigr),
\]
with initial condition given by the Stieltjes transform of $C_0$ (assuming a spectral density independent of $N_f$).
Upon ordering the $\omega_n$ along with increasing values of $\tau_n$, the solution finally reads
\begin{equation}\label{eq:MP2}
g(\rho,z) = g\Bigl(0,z+\rho\int_0^\infty  \frac{\tau d\sigma(\tau)}{1+\tau g(\rho,z)}\Bigr).
\end{equation}
Let us see how this solution can be recovered with the Dyson equation, based on Feynman diagrams expansion of the resolvent $G\n$.
We detail this formalism as it can be used conveniently also to compute
more complex quantities as train or test errors, i.e. product of matrices involving the resolvent multiple times.
The starting point is the expansion
\begin{equation}\label{eq:Gexpansion}
G\n(z) = G_0(z) +\sum_{p=1}^\infty(-1)^p G_0(z)\Bigl[\sum_{n=1}^N \omega_n\omega_n^t\Bigr]^p G_0(z),
\end{equation}
with $G_0(z)$ the resolvent of $C_0$.
When developing the terms $\Bigl[\sum_{n=1}^N \omega_n\omega_n^t\Bigr]^p$, averaging over the $\omega_n$ and taking the limit $N\to\infty$ at fixed $\rho = N/N_f$
we get a summation over terms, each one corresponding to a certain power $p$ in $\tau$ and $q$ in $\rho$, and organized topologically according to the Feynman diagram representation,
each one coming with a factor $\rho^q\tau^p$.
While $p$ is the number of terms $\omega_n\omega_n^t$ counted with their multiplicity  taken into account in the product, $q$ is the number of distinct terms $\omega_n\omega_n^t$ w.r.t. $n$.    
\begin{figure}[ht]
\centerline{\resizebox{1.0\textwidth}{!}{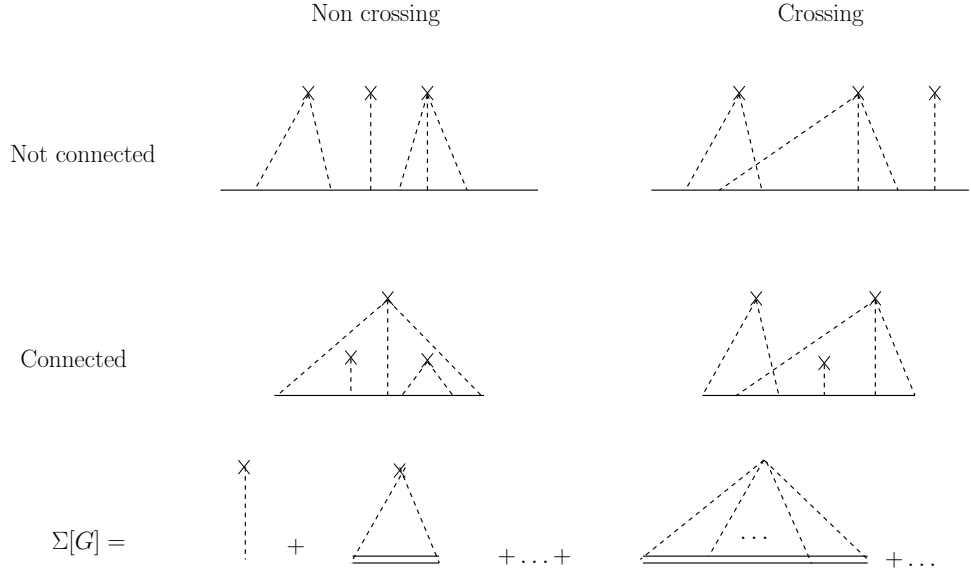}}
\caption{Examples of Feynman Diagrams corresponding to a contribution proportional to $\rho^3\tau^6$.
  The plain line corresponds to the propagator line and corresponds to insertion of the free propagator $G_0(z)$ between two successive interaction lines (dotted lines) or edges.
  The top left one for instance is  
  $-\rho^3{\mathbb E}\Bigl[G_0\omega_1\omega_1^tG_0\omega_1\omega_1^tG_0\omega_2\omega_2^tG_0\omega_3\omega_3^tG_0\omega_3\omega_3^tG_0\omega_3\omega_3^tG_0\Bigr]$,
  and is both non-crossing and not connected. The double plain line corresponds to use the full propagator instead in the expansion of the free energy. \label{fig:FeynmanDiag}}
\end{figure}
These diagrams can be classified in four global topological classes, depending on whether at least two interaction lines cross, and if the diagram can
be cut into 2 distinct diagrams by just cutting the propagator
line (see Figure~\ref{fig:FeynmanDiag}). The connected diagrams are actually not directly present in the expansion~(\ref{eq:Gexpansion}) because they are amputated from their external
propagator lines. They are building blocks of the complete expansion, similar in a way to what cumulants are for moments of a probability distribution. Defining the so-called self energy
$\Sigma_\rho$, in the field theory formalism,  as the sum over connected diagrams allows one to write the Dyson equation for the full propagator:
\[
G_\rho(z) = G_0(z) - \rho G_0(z)\Sigma_\rho G_\rho(z),
\]
i.e. the limit of $G\n$ in thermodynamic limit. 
In some cases the contribution of crossing diagrams becomes negligible in the asymptotic limit, and this corresponds precisely to the free probability setting~\cite{voiculescu1991limit,mingo2017free}.
Then the expansion simplifies considerably. The first point 
is that all factors of the form $\omega^t G_0\omega$ become deterministic and concentrated at its mean value at large $N$:
\[
\lim_{N_f\to\infty}\omega^tG_0(z)\omega = \lim_{N_f\to\infty} \Tr\Bigl[G_0(z)\omega\omega^t\Bigr] = \tau \lim_{N_f\to\infty} \frac{1}{N_f}\Tr\Bigl[G_0(z)\Bigr] = \tau g_0(z), 
\]
and the same remains valid when $G_0(z)$ is replaced by $G_\rho(z)$, as already stated in~\ref{eq:c1}. 
Then a simple combinatorial argument shows that the self-energy can be expressed in closed form function of the full propagator $G_\rho(z)$
as a sum of connected diagrams indicated on Figure~\ref{fig:FeynmanDiag}.
Then we have $\Sigma_\rho = \Sigma_\rho[G_\rho]$, with
\begin{equation}
\Sigma_\rho[G] =  N_f\sum_{n=1}^\infty (-1)^{n-1}{\mathbb E}\Bigl[\omega \bigl(\omega^t G\omega\bigr)^{n-1}\omega^t\Bigr]
= N_f{\mathbb E}\Bigl[\frac{\omega \omega^t}{1 + \omega^t G\omega}\Bigr],\label{eq:SelfEn}
\end{equation}
(the $N_f$ showing up comes from writing $\rho$ instead of $N$ in the previous equation).
Hence we get $G_\rho$ in a self-consistent form as 
\begin{equation}
G_\rho(z) = G_0(z) - \rho G_0(z)\Sigma_\rho\bigl[G_\rho(z)\bigr]G_\rho(z)\label{eq:Dyson},
\end{equation}
which leads to
\begin{equation}\label{eq:Gz}
G_\rho(z) = \Bigl(\I + \rho G_0(z)\Sigma_\rho\bigl[G_\rho(z)\bigr]\Bigr)^{-1}G_0(z).
\end{equation}
Note that $\Sigma(z)$ is noted ${\mathcal R}(z)$ and called the ${\mathcal R}$-transform in free probabilities~\cite{mingo2017free}.
Using~(\ref{eq:c1}) and the fact that the distribution of $\omega$ is isotropic
we obtain
\[
G_\rho(z) = G_0(z) - \rho G_0(z)\int_0^\infty d\sigma(\tau)\frac{\tau}{1+\tau g(\rho,z)} G_\rho(z).
\]
To recover (general)  Marcenko-Pastur solution we re-arrange this as
\[
g(\rho,z) = \Tr\left[\Bigl(1+\rho \int_0^\infty \frac{\tau d\sigma(\tau)}{1+\tau g(\rho,z)} G_0(z) \Bigr)^{-1} G_0(z)\right]
\]
after taking the trace.
Then thanks to the identity
\[
\Tr\bigl[G_0(z)^{n+1}\bigr] = \frac{(-1)^n}{n!}\frac{\partial^n}{\partial z^n}g(0,z),
\]
and formally identifying the Taylor expansion we arrive at~(\ref{eq:MP2}).

\section{Solution of the Dyson equation in planar approximation}\label{app:TrainTestproof}
We can now map the preceding equation to the specific problem under consideration to justify equations~(\ref{eq:Rasymp},\ref{eq:Etrainasymp}).
This justification will remains at a non-rigorous level, even though existing tools of RMT, like deterministic equivalents~\cite{hachem2007deterministic},
as those discussed and used in~\cite{Liu2020Ridge} on the same problem but assuming an isotropic signal,  would possibly
also work here to provide rigorous proofs.

At finite $N$ and $N_f$ our propagator is
\[
G\n = \bigl(\I+\alpha C\n)^{-1}.
\]
Therefore we have $z=1$, $G_0(z) = \I$ and
\[
\sum_{n=1}^N \omega_n^{(\sn)}\omega_n^{(\sn),t} = \alpha C\n,
\]
where the superscript $N$ as been added to underline the fact that the $\omega_n$ have a $1/\sqrt{N}$ scaling with $N$ at fixed $N_f$.
We are interested to give a meaning to the limit of this operator in the thermodynamic limit
\[
G_\rho = \lim_{N,N_f\to\infty\atop N/N_f=\rho} G\n.
\]
At finite $N_f$ we assume also that there is a prior $C$ to $C\n$, the population matrix, obtained by letting $N\to\infty$, so we have
\begin{equation}\label{eq:Eomomt}
{\mathbb E}\bigl(\omega^{(\sn)}\omega^{(\sn),t}\bigr) = \frac{\alpha C}{N}.
\end{equation}
Let us denote by $\omega_a\n$ the components of $\omega\n$ on the eigenvectors of $C$:
\[
\omega\n = \sum_{a=1}^{N_f}\omega_a\n \bu_a.
\]
We have
\begin{equation}\label{eq:moments}
{\mathbb E}\bigl[\omega_a\n\omega_b\n\bigr] = \delta_{a,b}\frac{\alpha c_a}{N}.
\end{equation}
The central point insuring the validity of the Marchenko-Pastur theory, hence justifying the planar approximation is
that the coefficient $\omega^{(\sn),t}G_\alpha\n\omega^{(\sn)}$ becomes deterministic in thermodynamic limits. If  
the random vector $\omega\n$ is assumed to have the moment constraint~(\ref{eq:moments}) 
and assuming typically some higher moments to be finite, then there exists some function $h(\rho,\alpha)$ such that the following holds:
\[
\lim_{N,N_f\to\infty} \omega\nt G\n\omega\n =  h(\rho,\alpha).
\]
As a result, combining this with (\ref{eq:SelfEn},\ref{eq:Eomomt}) we get  
\begin{equation}\label{eq:selfEnergy}
\Sigma_\rho[G_\rho] = \frac{\alpha C}{\rho\bigl[1+h(\rho,\alpha)\bigr]},
\end{equation}
therefore $G_\rho$, which expression in terms of the self-energy reads
\[
G_\rho = \bigl(\I+\rho\Sigma_\rho[G_\rho]\bigr)^{-1},
\]
commutes with $C$, being self-consistently defined as
\begin{equation}\label{eq:propagator}
G_\rho = \Bigl(\I+\frac{\alpha C}{\bigl[1+h(\rho,\alpha)\bigr]}\Bigr)^{-1},
\end{equation}
Note that these expressions are obtained by averaging, and we used them by convenience, but in a rigorous perspective we should instead use
deterministic equivalents~\cite{hachem2007deterministic} as is done in~\cite{Liu2020Ridge}. 
Nevertheless, this leads us to obtain a spectral decomposition of the Stieltjes transform of $C\n$ along the modes of $C$ in thermodynamic limit.
First define $g(x,\rho,\alpha)$ as
\[
\nu_\infty(x) g(x,\rho,\alpha) = \lim_{N,N_f\to\infty\atop N/N_f=\rho} \bu_a^t G\n \bu_a \delta(x-c_a),
\]
such that
\[
g(\rho,\alpha) \egaldef  \lim_{N,N_f\to\infty\atop N/N_f=\rho} \frac{1}{N_f}\Tr\bigl[G_\alpha\n\bigr] = \int \nu_\infty(dx) g(x,\rho,\alpha).
\]
Then we get 
\begin{align*}
  h(\rho,\alpha) &= \lim_{N,N_f\to\infty\atop N/N_f=\rho} \Tr\Bigl[G\n\omega\n\omega\nt\Bigr]\\[0.2cm]
  &= \lim_{N,N_f\to\infty\atop N/N_f=\rho} \frac{\alpha}{N}\Tr\Bigl[G\n C\Bigr]\\[0.2cm]
  &= \frac{\alpha}{\rho}\int \nu_\infty(dx) x g(x,\rho,\alpha)
\end{align*}
which is a special case of the Ledoit-P\'echet formula~\cite{ledoit2011eigenvectors},
along with the self-consistent equation 
\[
h(\rho,\alpha) = \lim_{N_f\to\infty}\frac{1}{N_f}\Tr\bigl[G\n C] = \frac{\alpha}{\rho}\int \nu_\infty(dx) x\Bigl(1+\frac{\alpha x}{1+h(\rho,\alpha)}\Bigr)^{-1}.
\]
In equation~(\ref{eq:propagator}) we see that the full propagator differs form the resolvent of $\alpha C$ by all
the contributions of correlations between data taken into the self energy. In this process the operator $\alpha C$
is replaced by $\rho\Sigma_\rho$ given in~(\ref{eq:selfEnergy}) or equivalently the bare coupling constant $\alpha$ is replaced by some effective coupling constant
\[
\Lambda(\rho,\alpha) = \frac{\alpha}{1+h(\rho,\alpha)},
\]
corresponding to the  dressing up of $\alpha$ with fluctuations in the traditional view-point of field theories.
This justifies the equation~(\ref{eq:Rasymp}) for the test-train error ratio,  where we see that ${\mathcal R}(\rho,\alpha) = 1+h(\rho,\alpha)$.
Concerning the train error~(\ref{eq:E_train}), to obtain its asymptotic counterpart~(\ref{eq:Etrainasymp}) we observe first that power of $G\n$ are obtained by derivation w.r.t. $\alpha$:
\[
\frac{\partial}{\partial\alpha}G\n = -\frac{1}{\alpha}G\n\bigl(\I-G\n\bigr).
\]
The fact that the expectation of $G_\rho$ is aligned with $C$ leads directly to~(\ref{eq:Etrainasymp}). In particular the first term expressed as a summation over
$\mu^\parallel(x)$ define by~(\ref{def:mupara}),
results from the fact that it is a self-averaging quantity as long as the transverse components $\tilde f_a = (F^+f)_a = \frac{f_a}{\sqrt{c_a}}$ on the modes of $C$ 
are evenly distributed as $\tilde f_a^2 = {\mathcal O}(1/\sqrt{N_f})$, i.e. without condensing onto a specific mode.

\end{document}